\documentstyle[preprint,eqsecnum,aps,floats,epsfig]{revtex}
\newif\iftightenlines\tightenlinesfalse
\tightenlines\tightenlinestrue

\begin{document}

\title{
\begin{flushright}
DO-TH 01/08\\ 
\end{flushright}
Neutrino Interactions In Oscillation Experiments
}

\author{E. A. Paschos\footnote{E-mail address:paschos@physik.uni-dortmund.de} and J. Y. Yu\footnote{E-mail address:yu@zylon.physik.uni-dortmund.de}}
\address{Institut f$\ddot{u}$r Physik, Universit$\ddot{a}$t Dortmund, D-44221 Dortmund, Germany}

\maketitle

\begin{abstract}

We calculate neutrino induced cross-sections relevant for oscillation experiments, including the $\tau$-lepton threshold for quasi-elastic, resonance and deep inelastic scattering. In addition to threshold effects, we include nuclear corrections for heavy targets which are moderate for quasi-elastic and large for single pion production. Nuclear effects for deep inelastic reactions are small. We present cross sections together with their nuclear corrections for various channels which are useful for interpreting the experimental results and for determining parameters of the neutrino sector. 
\\
\\
\end{abstract}

\begin{flushleft}

\vspace{6cm}

\end{flushleft}

\pagebreak

\section{INTRODUCTION}

Oscillation experiments \cite{Fukuda,Casper,Hirata} provide evidence for non-vanishing neutrino masses. Prominent among them is the reduction of the flux in atmospheric muon neutrinos and in solar neutrinos. To measure precisely the parameters $\delta m^2\simeq 10^{-2} - 10^{-3} \,\rm {eV^2}$ and $\sin^22\theta$, as well as to better understand the neutrino oscillation there are Long Baseline (LBL) experiments, like K2K \cite{K2k}, JHF-Kamioka \cite{Itow}\footnote{QE and resonances reactions are important at the K2K and JHF-Kamioka experiments because of the low neutrino energy ($E_\nu\simeq$ 1 GeV), see \protect\cite{Itow,Hall}. Since the beam energy is less than the threshold for the production of $\tau$-leptons (3.5 GeV) we do not consider them here.}, MINOS\cite{Minos}, OPERA\cite{Opera} and ICANOE\cite{Icanoe}, under construction and others being planned or running. The interest of LBL experiments lies mainly on the $\nu_\mu\rightarrow\nu_{\tau}$ channel, since it is already known from the CHOOZ experiment \cite{Apollonio}, that the possibility of $\nu_{\mu}\rightarrow \nu_e$ oscillation is very much supressed. In the LBL experiments several reactions contribute to neutrino nucleon reactions, namely deep inelastic scattering (DIS), quasi-elastic (QE) and resonance (RES) reactions. The latter two are low energy reactions, which are however still significant in the kinematic range of the new experiments with neutrino energies $E_\nu\simeq 4 - 30$ GeV. Therefore, these contributions should be included in the theoretical description of charged current reactions for $\tau$ appearance
\begin{equation}
\nu_\tau+ N \rightarrow \tau^- + X,
\end{equation}
\noindent
with $N = p,n$. In addition, we think it is useful to look at the various CC channels of $\nu_\tau$, searching for additional signatures which will help to confirm the reactions, since the expected number of $\tau$-events will be small. For this reason, we calculate the total cross sections and the number of events, $N_\tau$, for the deep inelastic, the quasi-elastic and the resonance  channels of CC reactions.
In addition, we consider NC reactions since they are important for establishing or eliminating oscillations into sterile neutrinos which are not completely excluded yet. 
In a previous articles we have already discussed the single pion production and the associated nuclear corrections \cite{Yu} of the resonances. Here we examine the DIS and QE channels including nuclear contributions as well. These results will be used to evaluate the number of events, $N_\tau$, for a heavy target like $_{26}{\rm Fe}^{56}$. For completeness, we also present anti-neutrino nucleon interactions of NC and CC reactions for QE and DIS with and without nuclear corrections.  \\
The paper is organized as follows: In section \ref{sec:two} we present the formalism and the evaluation of the charged and neutral current total cross section for deep inelastic scattering, quasi-elastic scattering and the resonance channels. In section \ref{sec:three} we explain theoretical aspects of the nuclear effects for the charged and neutral current total cross sections. Then we summarize our results for all the type of reactions including nuclear corrections. In section \ref{sec:four} we give the number of events, $N_\tau$, with and without nuclear corrections for CC and NC channels for the OPERA experiment. Several conclusions and their importance for the experiments are included in section \ref{sec:five}. 
 The results of this article should be useful for LBL experiments \cite{K2k,Itow,Minos,Opera,Icanoe} and those being discussed for the neutrino factory \cite{Alb}.
 
\section{General Formalism}\label{sec:two}

In this section we explain the main equations and the form factors used to calculate the cross section for DIS, QE and RES in (anti-)neutrino-nucleon interactions. Although the main contribution  comes from DIS, we treat also QE and RES because their contribution is still important in the energy region for LBL experiments. Since we have already discussed RES in our previous work \cite{Yu}, we outline mostly the calculations for DIS and QE scattering.  

\subsection{Deep Inelastic Scattering}

In this section we present the equations for the cross sections of deep inelastic scattering in $\nu(\bar{\nu})$-nucleon interactions. The CC channels are given by the following equations: 
\begin{eqnarray}
\nu_l(\bar{\nu}_l)(k_1) + N(p_1) \rightarrow l^{-}(l^{+})(k_2) + X(p_2).
\end{eqnarray}
The NC channels are:
\begin{eqnarray}
\nu_l(\bar{\nu}_l)(k_1) + N(p_1) \rightarrow \nu_l(\bar{\nu}_l)(k_2) + X(p_2)
\end{eqnarray}
where $N$ is a nucleon and $l = \mu, \tau$ and $X$ is the system of the outgoing hadrons. The double-differential cross section ${\rm d}\sigma/{\rm d}x{\rm d}y$ is:
\begin{equation}
\frac{{\rm d}\sigma}{{\rm d}x{\rm d}y} =\frac{G_F^2 y}{16 \pi}{\kappa^2 L^{\mu\nu} W_{\mu\nu}},
\end{equation}
where $\kappa = \frac{M_W^2}{Q^2+M_W^2}$ for the CC case, $G_F $ is the Fermi constant, $M_W$ is the W-boson mass and $y = \frac{\nu}{E_\nu}$. In the case of NC we have $\kappa = \frac{M_Z^2}{Q^2+M_Z^2}$ with $M_Z$ the Z-boson mass.\\
The leptonic tensor $ L^{\mu\nu}$ is:
\begin{equation}
L^{\mu\nu} = 2 {\rm Tr}[(\rlap/k_2+m_l) \gamma^\mu (1-\gamma_5) \rlap/k_1 \gamma^{\nu}],
\end{equation}
with the lepton mass $m_l$ for CC case, denoting $m_\tau$ or $m_\mu$. For the NC case we have $m_l = 0$.  
From Ref.\cite{Derman} the general hadronic tensor $W_{\mu\nu}$ is defined by:
\begin{eqnarray}
W_{\mu\nu}& = &- g_{\mu\nu} F_1(x,Q^2) + \frac{p_{1\mu}p_{1\nu}}{{p_1\cdot q}} F_              2(x,Q^2) - i \epsilon_{\mu\nu\rho\sigma} \frac{p_1^{\rho}                      q^{\sigma}}{2 p_1\cdot q} F_3(x,Q^2) + \nonumber \\
          &   & \frac{q_{\mu} q_{\nu}}{p_1\cdot q} F_4(x,Q^2) +
                (p_{1\mu}q_{\nu} + p_{1\nu}q_{\mu}) F_5(x,Q^2).
\end{eqnarray}
Here $\epsilon_{\mu\nu\rho\sigma}$ is the total antisymmetric tensor with $\epsilon_{0123}$ = + 1. $F_i (i =1...5)$ are the structure functions in neutrino-nucleon deep inelastic scattering. The differential cross section in the case of $m_l \neq 0$ is:
\begin{eqnarray}
\frac{{\rm d}\sigma^{\nu,\bar{\nu}}}{{\rm d}x{\rm d}y}& = &\frac{G_F^2 M E_\nu}{\pi} 
           \bigg [y\Big(xy + \frac{m_l^2}{2 E_\nu M}\Big)F_1 + 
            \Big(1-y -\frac{Mxy}{2 E_\nu} - \frac{m_l^2}{4 E_\nu^2}\Big) F_2 \pm \nonumber\\
      &   & \Big(xy(1-\frac{y}{2})-y\frac{m_l^2}{4 M E_\nu}\Big) F_3 + 
            \Big(x y \frac{m_l^2}{2 M E_\nu} + \frac{m_l^4}{4 M^2 E_\nu^2}\Big) F_4 -
            \frac{m_l^2}{2 M E_\nu} F_5\bigg],
\end{eqnarray}
where $x = \frac{Q^2}{2 M \nu}$ with $\nu = E_\nu-E_l$, $Q^2 = 2 M E_\nu x y$, $M$ the nucleon mass and the $\pm F_3$ signs correspond to the $\nu(\bar{\nu})$-nucleon scattering. 

To obtain the structure functions for the proton and the neutron for charged and neutral channels we used the quark parton model (QPM). We treated the proton and the neutron separately in order to account for non-isoscalar targets. The structure functions of the charged current reactions for the $\nu(\bar{\nu})$-proton scattering above the threshold for charm production for the proton are:
\begin{eqnarray}
F_2^{CC}(\nu p)& = & 2 x [d + s + \bar{u} +\bar{c}]\\\nonumber
x F_3^{CC}(\nu p)& = & 2 x [d + s - \bar{u} -\bar{c}]\\\nonumber
F_2^{CC}(\bar{\nu} p)& = & 2 x [u + c + \bar{d} +\bar{s}]\\\nonumber
x F_3^{CC}(\bar{\nu} p)& = & 2 x [u + c - \bar{d} -\bar{s}]
\end{eqnarray}
and for the neutron
\begin{eqnarray}
F_2^{CC}(\nu n) & = & 2 x [u + s + \bar{d} +\bar{c}]\\\nonumber
x F_3^{CC}(\nu n) & = & 2 x [u + s - \bar{d} -\bar{c}]\\\nonumber
F_2^{CC}(\bar{\nu} n) & = & 2 x [d + c + \bar{u} +\bar{s}]\\\nonumber
x F_3^{CC}(\bar{\nu} n) & = & 2 x [d + c - \bar{u} -\bar{s}]. 
\end{eqnarray}

Below the threshold for charm production the corresponding structure functions for the proton are:
\begin{eqnarray}
F_2^{CC}(\nu p)& = & 2 x [d\cos^2\theta_c + s\sin^2\theta_c + \bar{u} +\bar{c}]\\ \nonumber
x F_3^{CC}(\nu p)& = & 2 x [d\cos^2\theta_c + s\sin^2\theta_c - \bar{u} -\bar{c}]\\ \nonumber
F_2^{CC}(\bar{\nu} p)& = & 2 x [u\cos^2\theta_c + c\sin^2\theta_c + \bar{d} +\bar{s}]\\ \nonumber
x F_3^{CC}(\bar{\nu} p)& = & 2 x [u\cos^2\theta_c + c\sin^2\theta_c - \bar{d} -\bar{s}]
\end{eqnarray}
and for the neutron 
\begin{eqnarray}
F_2^{CC}(\nu n) & = & 2 x [u\cos^2\theta_c + s\sin^2\theta_c + \bar{d} +\bar{c}]\\ \nonumber
x F_3^{CC}(\nu n) & = & 2 x [u\cos^2\theta_c + s\sin^2\theta_c -\bar{d} -\bar{c}]\\ \nonumber
F_2^{CC}(\bar{\nu} n) & = & 2 x [d\cos^2\theta_c + c\sin^2\theta_c + \bar{u} +\bar{s}]\\ \nonumber
x F_3^{CC}(\bar{\nu} n) & = & 2 x [d\cos^2\theta_c + c\sin^2\theta_c -\bar{u} -\bar{s}],
\end{eqnarray}  
with Cabibbo angle $\cos\theta_c = 0.9755$ \cite{Particle Data}.
Notice that the contribution of the threshold effect for charm production is small ($\simeq$ 5$\%$) and negligible.

The neutral current reactions for the $\nu(\bar{\nu})$-proton scattering depends on:
\begin{eqnarray}
F_2^{NC}(\nu p,\bar{\nu} p)& = & 2 x ((g_L^2+g_R^2) [u + c + \bar{u} +\bar{c}]+(g_L^{\prime 2}+g_R^{\prime 2}) [d + s + \bar{d} +\bar{s}])\\\nonumber
x F_3^{NC}(\nu p,\bar{\nu} p) & = &  2 x ((g_L^2-g_R^2) [u + c -\bar{u} -\bar{c}]+(g_L^{\prime 2}-g_R^{\prime 2}) [d + s -\bar{d} -\bar{s}])\nonumber
\end{eqnarray}
and for the neutron
\begin{eqnarray}
F_2^{NC}(\nu n, \bar{\nu} n) & = & 2 x ((g_L^2+g_R^2) [d + c + \bar{d} +\bar{c}]+(g_L^{\prime 2}+g_R^{\prime 2}) [u + s + \bar{u} +\bar{s}])\\\nonumber
x F_3^{NC}(\nu n, \bar{\nu} n) &  = & 2 x ((g_L^2-g_R^2) [d + c -\bar{d} -\bar{c}]+(g_L^{\prime 2}-g_R^{\prime 2}) [u + s -\bar{u} - \bar{s}]),\nonumber
\end{eqnarray}  
where $g_{L} = \frac{1}{2} -\frac{2}{3} \sin^2 \theta_W, \,g_{R} = -\frac{2}{3} \sin^2 \theta_W$ and $g_L^\prime = - \frac{1}{2} + \frac{1}{3} \sin^2 \theta_W, \,g_R^\prime = \frac{1}{3} \sin^2 \theta_W$ with the Weinberg angle $\sin^2 \theta_W = 0.23117$ \cite{Particle Data}.
To calculate the cross section we integrate $\frac{{\rm d}\sigma}{{\rm d}x{\rm d}y}$ for the muon case in the range $0\leq x\leq 1$ and $0\leq y\leq 1$. For the tau case we use the limits \cite{Albright}:
\begin{eqnarray}
& & \frac{m_\tau^2}{(2M E_\nu -\sqrt{2} m_\tau M)}\leq x \leq 1 \\ \nonumber
& & a - b \leq x \leq  a + b,\nonumber
\end{eqnarray}
where $a$ and $b$ are defined the following way:
\begin{eqnarray}
a & = & \frac{1 - m_\tau^2(\frac{1}{2 M E_\nu x} + \frac{1}{2E_\nu^2})}{2(1 +  \frac{Mx}{2E_\nu})}\\ \nonumber
b & = & \frac{\sqrt{(1 - \frac{m_\tau^2}{2ME_\nu x})^2 - \frac{m_\tau^2}{E_\nu^2}}}{2(1+\frac{Mx}{2E_\nu})}.\nonumber
\end{eqnarray}
For the quark distributions we use the CTEQ5 leading order (LO) parton distributions \cite{Lai}. 
The Callan-Gross relation relates $F_2$ to $F_1$ :
\begin{equation}
2 x F_1 = F_2.
\end{equation}
Furthermore, we use for $F_4$ and $F_5$ the Albright-Jarlskog relations \cite{Albright}: 
\begin{eqnarray}
F_4 & = & 0\\
x F_5 & = & F_2.
\end{eqnarray}

\subsection{Quasi-Elastic  Neutrino Scattering}

Following Ref.\cite{Llewellyn} we calculated the charged current and neutral current channels of $\nu(\bar{\nu}) N$ reactions:
\begin{eqnarray}
\nu_l(\bar{\nu}_l)(k_1) + N(p_1) &\rightarrow & l^{-}(l^{+})(k_2) + N(p_2) \qquad(\text{CC}) \\
\nu_l(\bar{\nu}_l)(k_1) + N(p_1) &\rightarrow & \nu_l(\bar{\nu}_l)(k_2) + N(p_2)\qquad(\text{NC}).
\end{eqnarray}
Notice that the neutral current reaction on neutron targets is in practice not measurable, even though its cross section is comparable to the proton reaction. The matrix element of the quasi-elastic reactions can be defined as follows:
\begin{equation}
{\em M}= \frac{ig^2\cos\theta_c}{4}\frac{g_{\mu\nu}}{q^2-M_w^2}\bar{u}(k_2)\gamma^\mu(1-\gamma_5)u(k_1)\bar{u}(p_2)\Gamma^\nu u(p_1).
\end{equation}
$\Gamma^\nu$ is given by:
\begin{eqnarray}
\Gamma^\nu & = & \gamma^\nu F_1^V(q^2)+i \sigma^{\nu\alpha} \frac{q_{\alpha} \xi F_2^V(q^2)}{2 M} + \nonumber \\         
 &   & \frac{q^\nu F_3^V(q^2)}{M} +\gamma^\nu \gamma_5 F_A(q^2) +
 \frac{q^\nu \gamma_5 F_p(q^2)}{M} + \frac{\gamma_5 (p_1+p_2)^\nu}{M} F_3^A(q^2),
\end{eqnarray}
where $F_i^V \,(i = 1,2,3),\, F_A,\, F_3^A,\, F_p$ are the weak form factors of the nucleon.
The form factors are in general complex, but general principles eliminate two of them and require the rest to be real.
First, $F_i^V, \, F_A,\, F_p$ and $F_3^A$ are real because of time reversal invariance. Second, $F_1^V,F_2^V,F_A$ and $F_p$ are real but $F_3^A$ and $F_3^V$ are imaginary because of charge symmetry. Thus $F_3^A = F_3^V  = 0$ (no second class currents) from  these two constraints. 
The conserved vector current (CVC) hypothesis establishes the following relations between the weak form factors and the electromagnetic form factors:
$F_1^V(q^2) = F_1^p(q^2) - F_1^n(q^2)$ and $\xi F_2^V(q^2) = \mu_p F_2^p(q^2)-\mu_n F_2^n(q^2)$ with $\xi = \mu_p-\mu_n = 3.706, \, k_p = \mu_p -1 = 1.793$ and $k_n = \mu_n = -1.913$. $k_p$ and $k_n$ are the anomalous magnetic moments of proton and neutron and $F_{1}^{p,n}$ and $F_{2}^{p,n}$ are the electromagnetic Dirac-Pauli isovector form factors of proton and neutron. $F_1^V(q^2)$ and $F_2^V(q^2)$ can be expressed in terms of the Sachs form factors:
\begin{eqnarray}
F_1^V(q^2) & = & \frac{G_E^V(q^2)-\frac{q^2}{4 M^2} G_M^V(q^2)}{1 - \frac{q^2}{4 M^2}}\\
\xi F_2^V(q^2) & = & \frac{G_M^V(q^2)- G_E^V(q^2)}{1 - \frac{q^2}{4 M^2}},
\end{eqnarray}
where
\begin{eqnarray}
G_E^V(q^2) & = & \frac{1}{(1 - \frac{q^2}{M_V^2})^{2}}\\
G_M^V(q^2) & = & \frac{1+\xi}{(1 - \frac{q^2}{M_V^2})^{2}}
\end{eqnarray}
 with a vector mass $M_V = 0.84$ GeV. The axial vector form factor is given by:
\begin{equation}
F_A(q^2) =\frac{F_A(0)}{(1- \frac{q^2}{M_A^2})^{2}} 
\end{equation}
with an axial vector mass $M_A = 1.0$ GeV and $F_A(q^2 = 0) = - 1.23$. A reasonable approximation for all $q^2$ is given by:
\begin{equation}
F_p(q^2) = 2 M^2 \frac{F_A(q^2)}{(M_\pi^2 - q^2)} 
\end{equation}
with the pion mass $M_\pi = 0.14$ GeV. 

For NC reactions, we replace the charged current form factors with the neutral current form factors. In the electroweak theory the charged current form factors are related to the neutral current form factors as follows:
\begin{eqnarray}
(F_1^V)^{NC}(q^2)   & = & \frac{1}{2} F_1^V(q^2) -2 \sin^2\theta_W F_1^p(q^2)\\
\xi (F_2^V)^{NC}(q^2)& = & \frac{1}{2} \xi F_2^V(q^2) -2 \sin^2\theta_W \,(\mu_p-1) F_2^p(q^2)\\
F_A^{NC}(q^2) & = &\frac{1}{2} F_A(q^2) \\
F_p^{NC}(q^2) & = & \frac{2 M^2 F_A^{NC}(q^2)}{(M_\pi^2 - q^2)},
\end{eqnarray}
where 
\begin{eqnarray}
F_1^N(q^2) & = & \frac{G_E^N(q^2)-\frac{q^2 G_M^N(q^2)}{4 M^2}}{1-\frac{q^2}{4 M^2}}, \quad\,
\mu_N F_2^N(q^2) = \frac{G_M^N(q^2)- G_E^N(q^2)}{1-\frac{q^2}{4 M^2}}, \\
G_E^N(q^2) & = &\frac{G_{E}^N(0)}{(1 - \frac{q^2}{M_V^2})^{2}},\quad\, 
G_M^N(q^2) = \frac{G_{M}^N(0)}{(1 -\frac{q^2}{M_V^2})^{2}}.
\end{eqnarray}
At $q^2$ = 0 the form factors are normalized by the following conditions: 
\begin{eqnarray}
G_E^p(0) = 1, \, G_E^n(0) = 0,\, G_M^N(0) = \mu_N,\,
F_A^{NC}(0) = -0.615.
\end{eqnarray}
After some standard but tedious algebra we arrive at the differential cross section:
\begin{eqnarray}
{{{\rm d}\sigma}\over{{\rm}d |q^2|}} & = &\frac{G_F^2 \cos^2\theta_c}{8 \pi E_\nu^2}\bigg[(F_1^V)^2 \frac{q^4 - 4 M^2 (m_l^2 - q^2) - m_l^4}{4 M^2} +\nonumber \\
 &  & (\xi F_2^V)^2 \frac{4 M^2 (q^4 - m_l^4) - q^4(m_l^2 - q^2)}{16 M^4} + (F_A^V)^2 \frac{q^4 + 4 M^2 (m_l^2 - q^2) - m_l^4}{4 M^2} - \nonumber \\
&  & (F_p)^2 \frac{m_l^2 q^2 (-q^2 + m_l^2)}{4 M^4} + F_1^V \xi F_2^V \frac{2 q^4 + q^2 m_l^2 + m_l^4}{2 M^2} - \nonumber \\
&  & F_A F_P \frac{m_l^2 (- q^2 + m_l^2)}{2 M^2} + F_A (F_1^V + \xi F_2^V) q^2 \frac{(s-u)}{M^2} + \nonumber \\
& & \Big((F_1^V)^2 -  \frac{(\xi F_2^V)^2 q^2}{4 M^2} + (F_A)^2\Big) \frac{(s-u)^2}{4 M^2}\bigg], 
\end{eqnarray} 
with $ s - u = 4 E_\nu M + q^2-m_l^2$. For the anti-neutrino nucleon reactions in QE channels we replace the term $F_A (F_1^V + \xi F_2^V)$ by $- F_A (F_1^V + \xi F_2^V)$.

\subsection{RES}

In a previous paper\cite{Yu} we discussed in detail the neutral current differential cross section for the production of resonances on various materials and included nuclear effects. In this article we include in addition the $\nu_\tau(\bar{\nu}_\tau)$-nucleon interactions of CC channels, like:
\begin{equation}
\nu_\tau(\bar{\nu}_\tau) + p \rightarrow \tau^- (\tau^+) + p +\pi^+(\pi^-)
\end{equation}
\begin{equation}
\nu_\tau(\bar{\nu}_\tau) + n \rightarrow \tau^- (\tau^+)+ n +\pi^+(\pi^-)
\end{equation}
\begin{equation}
\nu_\tau(\bar{\nu}_\tau) + n(p)\rightarrow \tau^-(\tau^+) + p(n) +\pi^0  
\end{equation}
where $m_\tau$ = 1.78 GeV and calculate the differential cross sections with respect to the pion energy as well as the total cross sections. We present the results for various incoming neutrino energies and include nuclear corrections for the different nuclei, using the same kinematics as in our previous article \cite{Yu}. 

\section{nuclear effects}\label{sec:three}

As mentioned already the heavy nuclei of the targets bring additional effects. We discuss them separately for the various reactions, considering $_{26}{\rm Fe}^{56}$ as a typical target. We investigate in this paper nuclear effects for the DIS and QE reactions. In addition, we extend the method of \cite{Yu} to the charged current channels of the resonances. Nuclear corrections for the neutral current case are contained in Ref. \cite{Yu}.

\subsection{DIS}

For the nuclear corrections of deep inelastic scattering we use two different sets of parton distributions, the $\chi^2$-analysis of Ref. \cite{Kum} as well as the  EKS98 parameterization \cite{Esk,Esk1}. Nuclear parton distributions are determined for the EKS98 parameterization at an initial scale $Q_0^2$ = 2.25 $\rm {GeV^2}$ and in the $x$ range $10^{-6} \leq x \leq 1$ through a DGLAP evolution, using the data from lepton-nucleus ($l$A) DIS and Drell-Yan (DY) measurements from proton-nucleus (pA) collisions with conservation of momentum and baryon number as constraints. Their nuclear modifications contain shadowing for $x\leq 0.1$, anti-shadowing for $0.1\leq x \leq 0.3$, EMC effect for $0.3\leq x \leq 0.7$ and Fermi motion for $x\rightarrow 1$. The first method bases on a $\chi^2$-analysis of the data, which was taken from deep inelastic electron and muon scattering, provides nuclear structure functions at the initial scale $Q_0^2$ = 1.0 $\rm {GeV^2}$ and the $x$ range $10^{-9} \leq x \leq 1$. The $\chi^2$-analysis does not contain the charm distribution. Their results are quite sensitive on the Bjorken variable $x$ and there is a slight difference between the $\chi^2$-analysis from Ref. \cite{Kum} and the EKS98 Parameterization \cite{Esk,Esk1}. However, the difference occurs just in the sea quark distribution and is only noticeable in the small $x$ region.      

\subsection{QE}

Important nuclear effects for the quasi-elastic scattering arise from the Pauli principle, rescattering and absorption of recoiling hadrons and from the Fermi motion. We use only the Pauli principle since it is the most important nuclear effect and neglect the other two, see \cite{Lov,Yao,Battistoni} for a detailed discussion. We calculated the Pauli factor of the quasi-elastic scattering according to Refs.\cite{Llewellyn,Bell,Singh}. We multiply for neutrons the Pauli factor $g = 1- N^{-1} D$ with the total cross section for QE where:
\begin{equation}
D =\left \{ \begin{array}{ll}
      Z & \mbox{for $ 2 x < u-v$}\\
      0.5 A \Big(1-\frac{3 x (u^2+v^2)}{4}+ \frac{x^3}{3}-\frac{3(u^2-v^2)^2}{32 x}\Big)  & \mbox{for $u-v <2 x <u+v$}\\
      0 & \mbox{for $2 x > u+v$}, 
\end{array} 
\right.
\end{equation}
where $ x =\frac{|\bf q|}{2 k_F}, u =(\frac{2 N}{A})^{\frac{1}{3}}$ and $v = (\frac{2 Z}{A})^{\frac{1}{3}}$. The Fermi momentum is taken $k_F$ = 1.36 ${\rm fm}^{-1}$ from Ref. \cite{Bohr}. $N, Z$  and $A$ are neutron, proton, nucleon number, respectively. The three-momentum transfer $|{\bf q}|$ is defined by $\frac{q^2}{2 M}\sqrt{1-\frac{4 M^2}{q^2}}.$ For protons we just replace $N$ by $Z$.   

\subsection {Results}

We present results for $\nu_\mu$ and $\nu_\tau$ induced reactions. All $\nu_\tau$ charged current reactions show an evident $\tau$-lepton threshold with the cross sections becoming large and noticeable for beam energies above $5 - 6$ GeV. The neutral current reactions do not show any threshold but they are smaller than the $\nu_\mu$ reactions by a factor of approximately ten.\\
The energy dependence is the second difference: The QE and RES total cross sections reach constant asymptotic values at high energies, while the DIS cross section rises linearly with energy. For this reason we plot for DIS the ratios $\sigma/E_\nu$ or $\sigma/(G_F^2 m_N E_\nu/\pi)$. Immediately above the threshold the DIS process dominates over the QE and RES, while below  $E_\nu < 5$ GeV  the sum of QE and resonance is approximately 50$\%$ of the total cross section. We discuss next each of the reactions separately.
 
\subsubsection{Deep Inelastic scattering}

\begin{figure}[b]
\centerline{\epsfig{figure=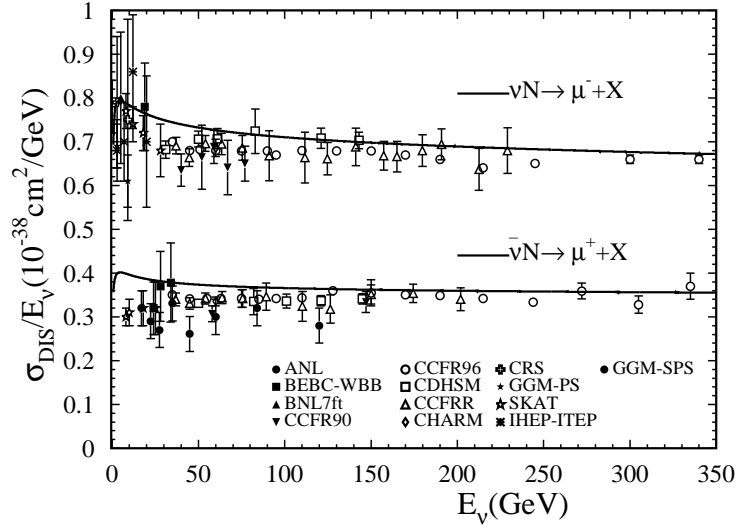,height=3.5in,angle=0}}
\caption{The cross section of DIS for the $\nu_\mu + N \rightarrow \mu^- + X$and $\bar{\nu}_\mu + N \rightarrow \mu^+ + X$ plotted versus the incoming neutrino energy for the isoscalar target with the normalization of $1/E_\nu$. 
The data points have been taken from ANL \protect\cite{Barish}, BEBC-WBB \protect\cite{Dolly}, BNL7ft \protect\cite{Baker}, CCFR90 \protect\cite{Auc}, CCFR96 \protect\cite{Seligman}, CCFRR \protect\cite{Mac}, CDHSW \protect\cite{Berge}, CHARM \protect\cite{All}, CRS \protect\cite{Bal}, GGM-PS \protect\cite{Cam}, GGM-SPS \protect\cite{Mor}, IHEP-ITEP \protect\cite{Vov}, SKAT \protect\cite{Bar}.}\label{dismu}
\end{figure}

\begin{figure}[b]
\centerline{\epsfig{figure=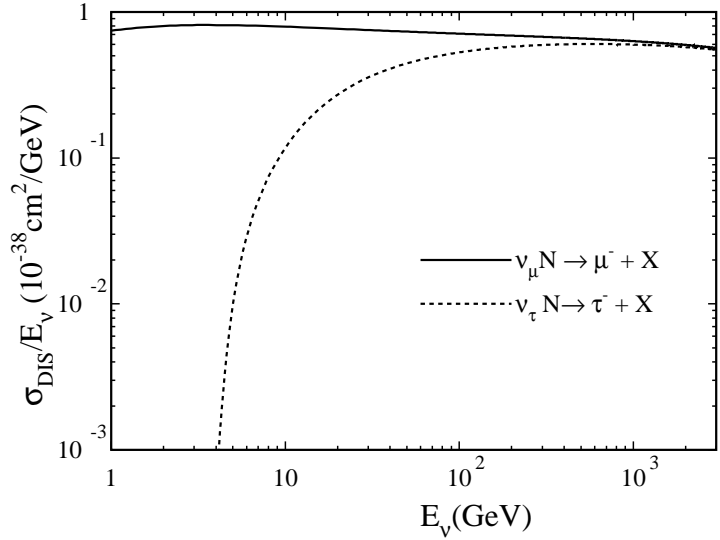,height=3.5in,angle=0}}
\caption{The cross section of DIS for $\nu_\tau + N\rightarrow \tau^- + X$ and $\nu_\mu + N\rightarrow \mu^- + X$ for the isoscalar target plotted versus the incoming neutrino energy with the normalization of $1/E_\nu$.} \label{totc}
\end{figure}

\begin{figure}[b]
\begin{minipage}[b]{.49\linewidth}
\centering\epsfig{file=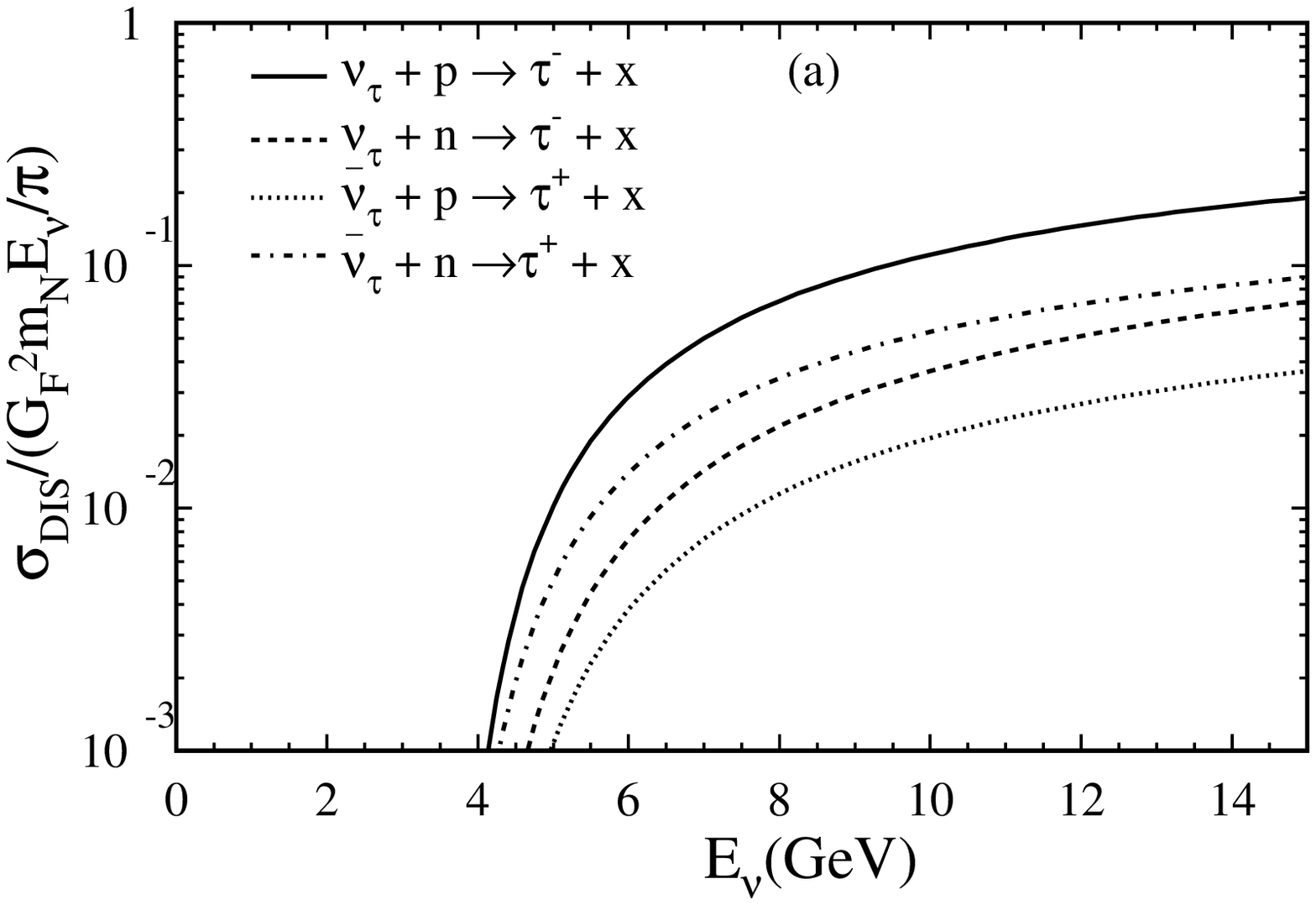,height=3.in,width=\linewidth}
\end{minipage}
\begin{minipage}[b]{.49\linewidth}
\centering\epsfig{file=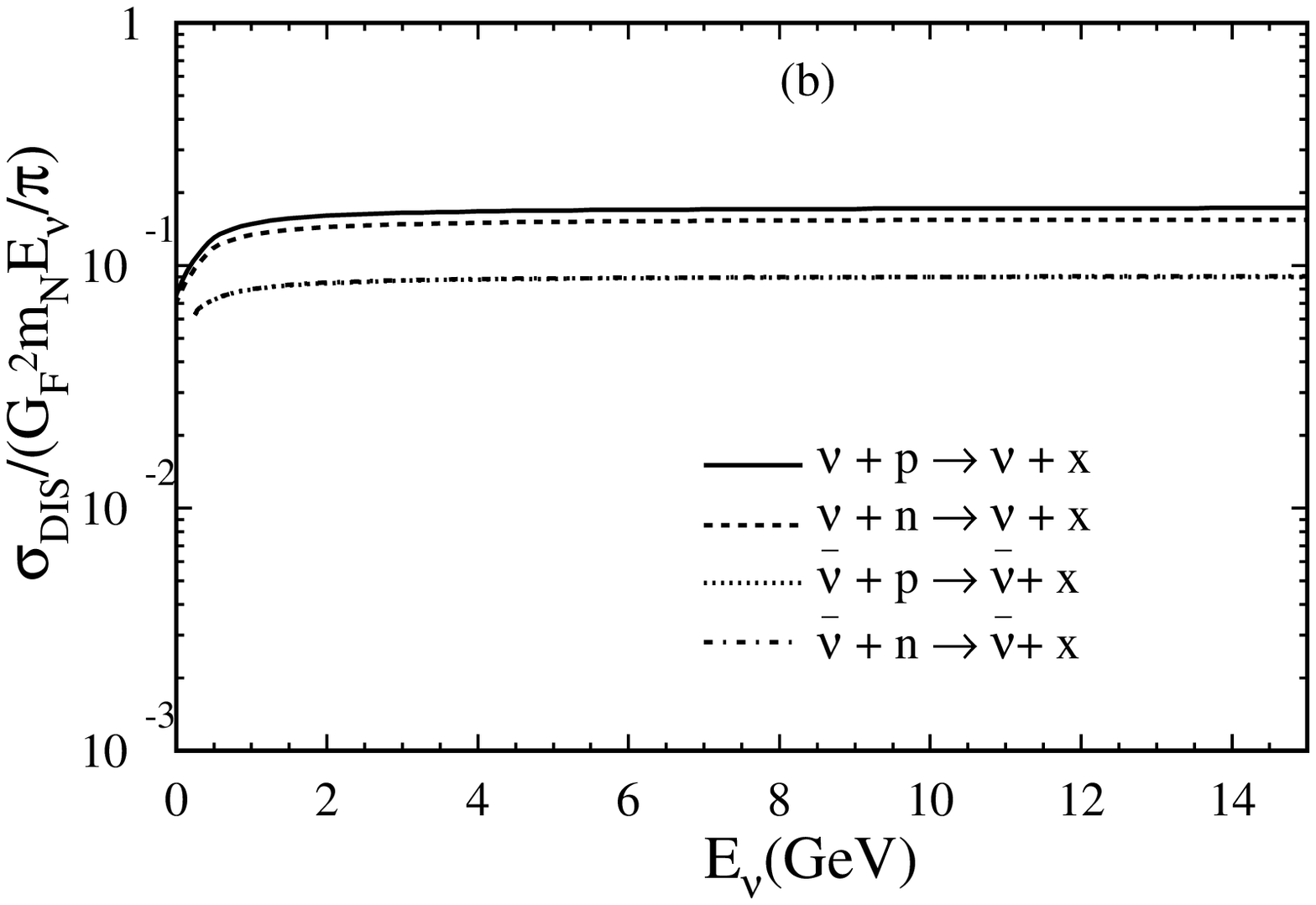,height=3.in,width=\linewidth}
\end{minipage}
\vspace{-1.0cm}
\caption{The cross section of DIS for (a) $\nu_\tau(\bar{\nu}_\tau) + N\rightarrow \tau^-(\tau^+) + X$ and (b) $\nu(\bar{\nu}) + N\rightarrow \nu(\bar{\nu}) + X$ plotted versus the incoming neutrino energy, normalized by $G_F^2 m_N E_\nu/\pi$.}\label{taucon}
\begin{minipage}[b]{.49\linewidth}
\centering\epsfig{file=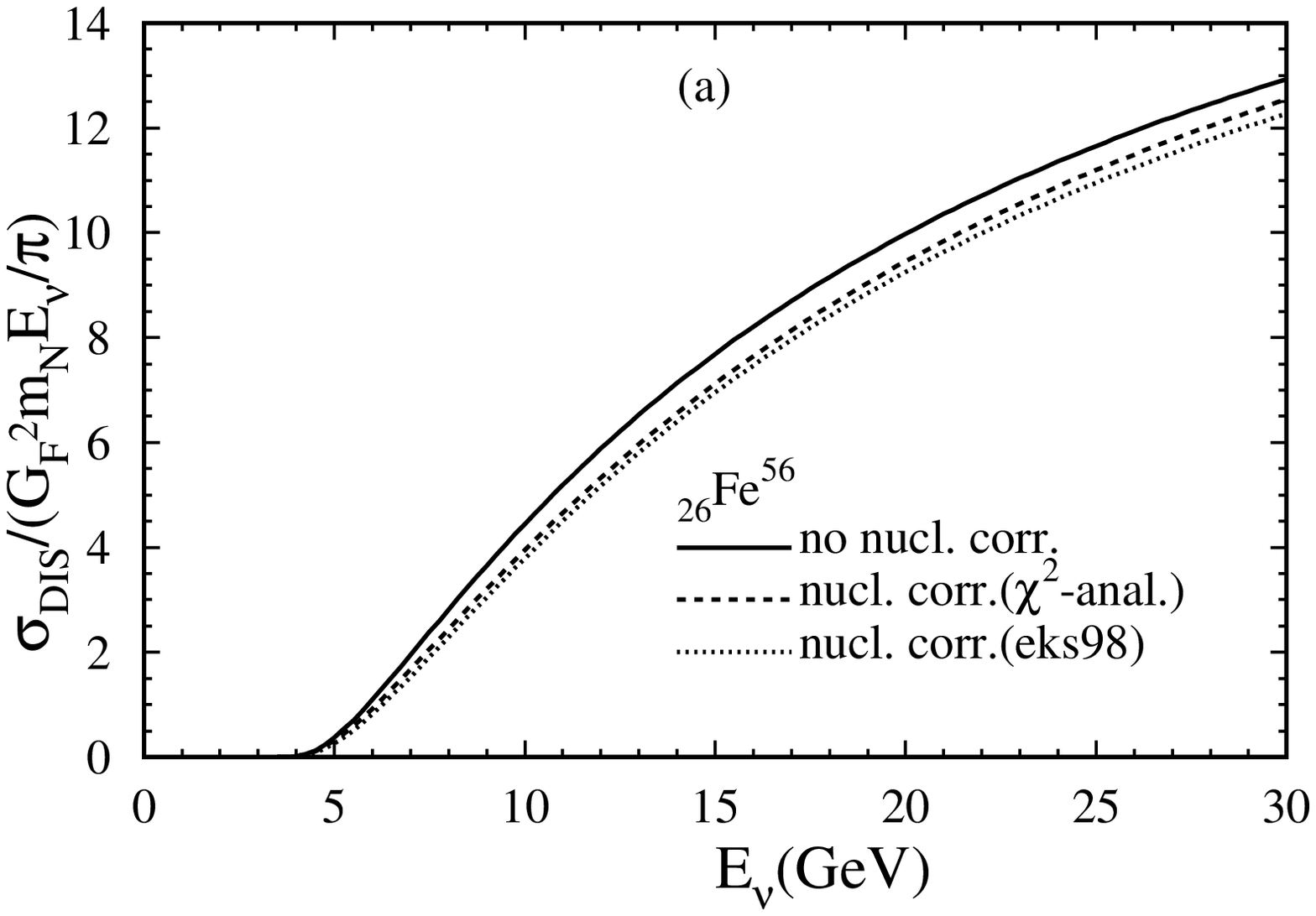,width=\linewidth}
\end{minipage}
\begin{minipage}[b]{.49\linewidth}
\centering\epsfig{file=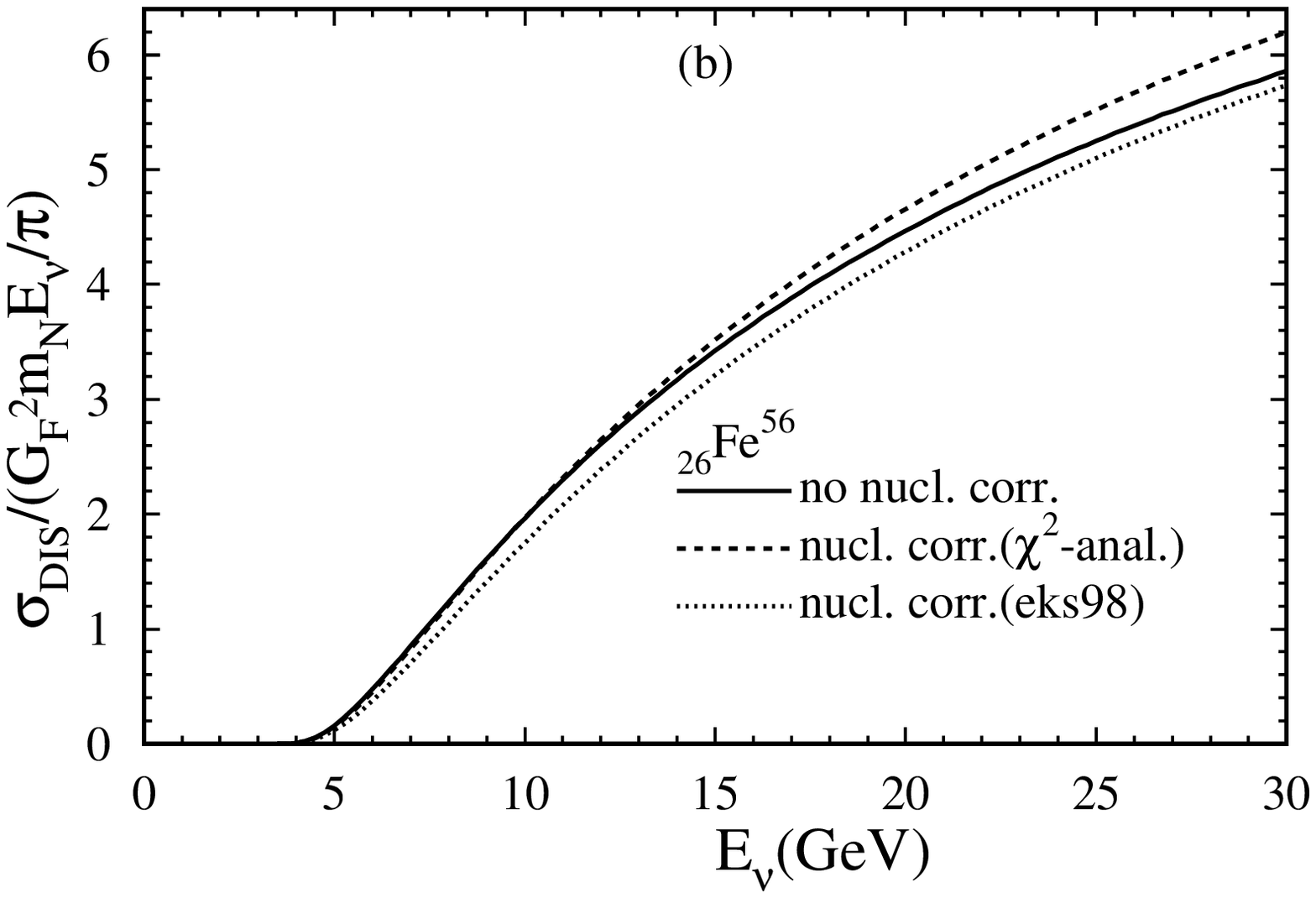,width=\linewidth}
\end{minipage}
\vspace{-1.0cm}
\caption{The cross section of (a) $\nu_\tau$ and $\bar{\nu}_\tau$ charged current reactions for DIS on iron targets versus the incoming neutrino energy, normalized by $G_F^2 m_N E_\nu/\pi$. The solid, dashed and dotted curves represent the cross section of DIS with and without nuclear corrections of $\chi^2$ analysis \protect\cite{Kum} and EKS98 parameterization \protect\cite{Esk,Esk1}.}\label{taunuc}

\begin{minipage}[b]{.49\linewidth}
\centering\epsfig{figure=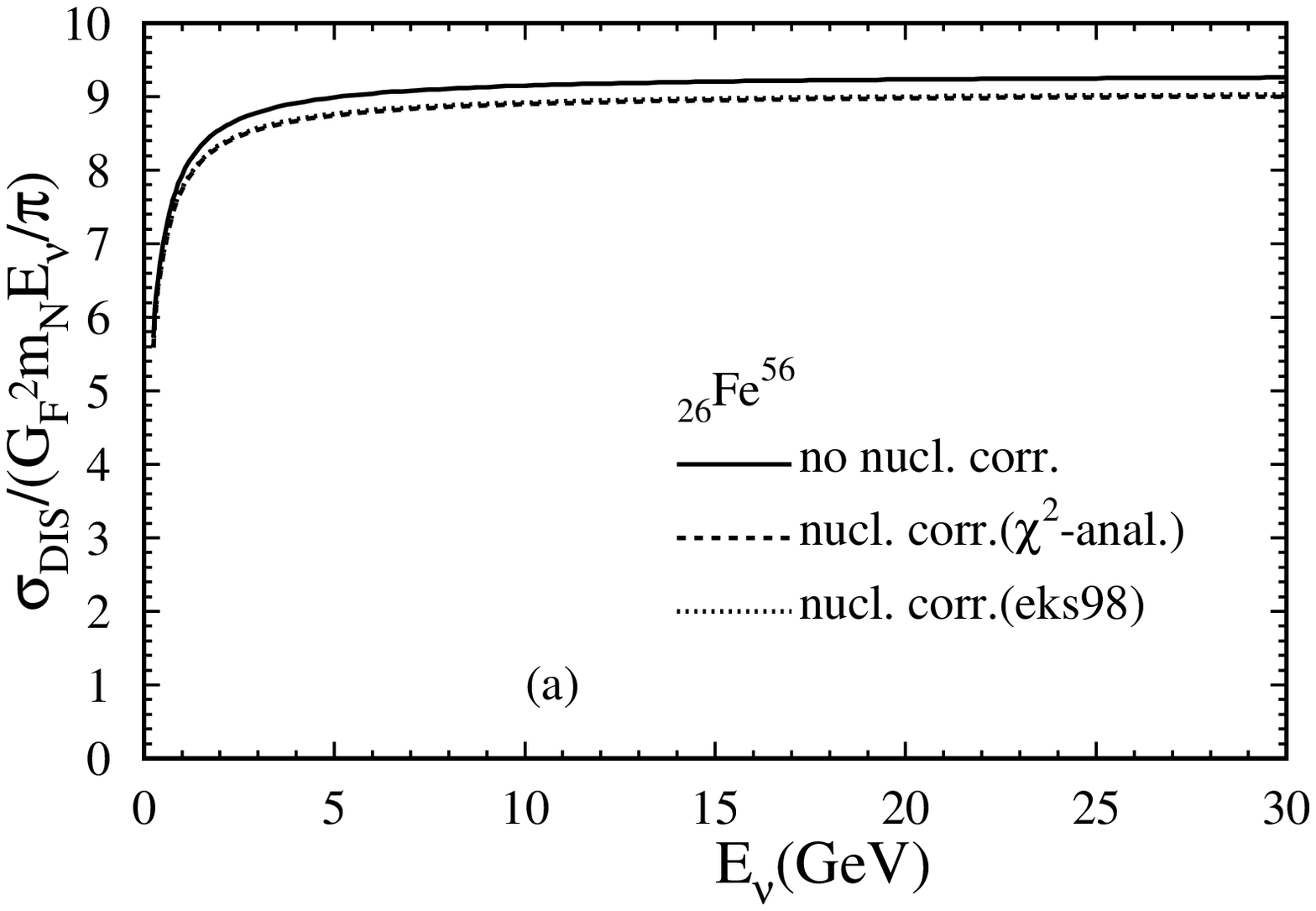,width=\linewidth}
\end{minipage}
\begin{minipage}[b]{.49\linewidth}
\centering\epsfig{file=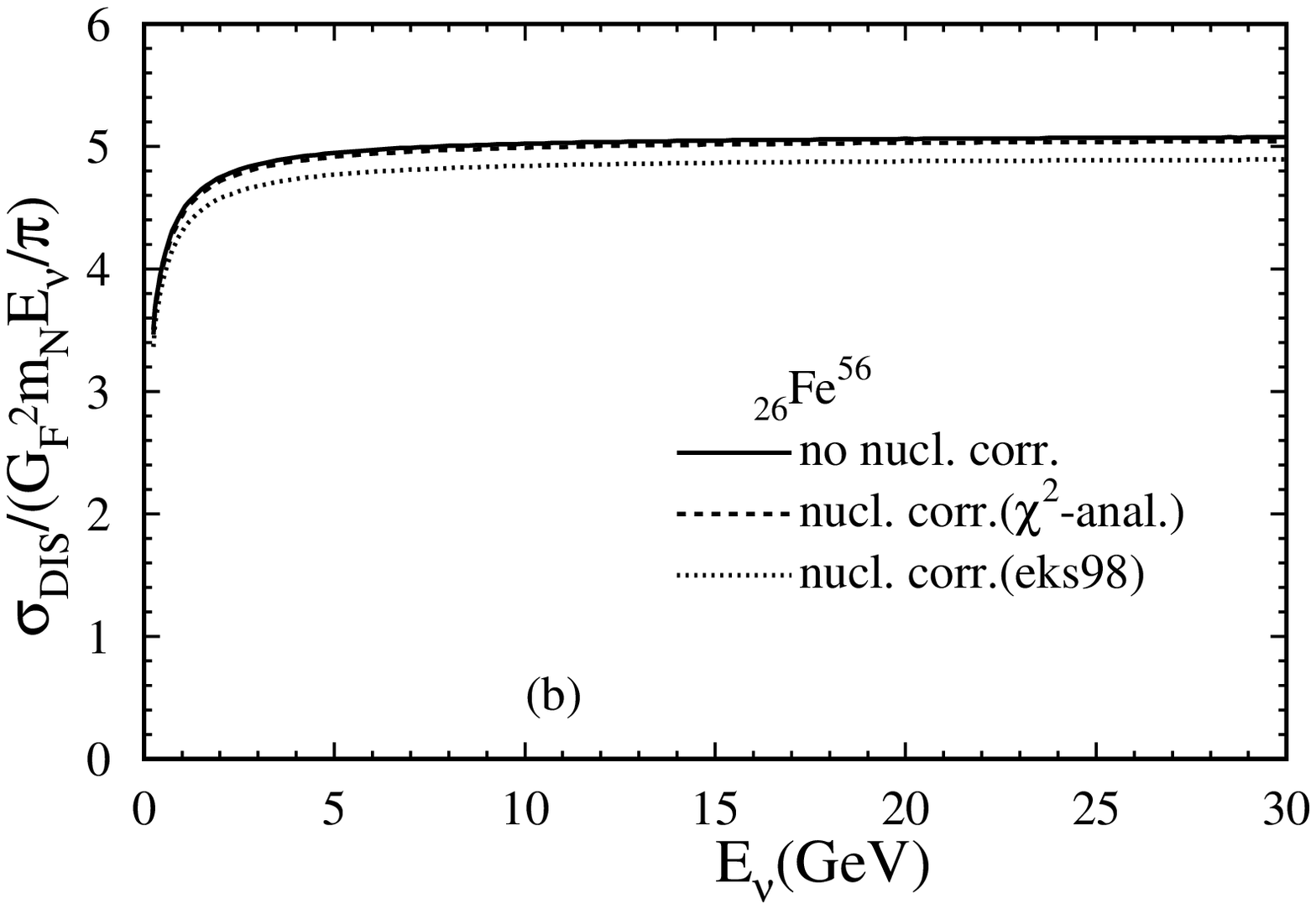,width=\linewidth}
\end{minipage}
\vspace{-1.0cm}
\caption{The cross section of (a) $\nu$ and (b) $\bar{\nu}$ neutral current reactions for DIS on iron targets versus the incoming neutrino energy, normalized by $G_F^2 m_N E_\nu/\pi$. The solid, dashed and dotted curves represent the cross section of DIS with and without nuclear corrections of $\chi^2$ analysis \protect\cite{Kum}  and EKS98 parameterization \protect\cite{Esk,Esk1}.}\label{ncnuc}
\end{figure}

In Fig.~\ref{dismu} we show the total cross sections for $\nu_\mu(\bar{\nu}_\mu) + N\rightarrow \mu^-(\mu^+) + X$ as a function of neutrino energy. More precisely, we plot the slopes vs energy. We see that the curves start with a constant slope and remain so up to 350 GeV. We also include the experimental data from various groups \cite{Barish,Dolly,Baker,Auc,Seligman,Mac,Berge,All,Bal,Cam,Mor,Vov,Bar} and the agreement is good. Fig.~\ref{totc} shows the reaction $\nu_\tau + N\rightarrow \tau^- + X$ where the threshold dependence from the mass of the $\tau$-lepton is now evident. For comparison we included the $\nu_\mu$ cross section.
 The $\nu_\mu$ and $\nu_\tau$ induced reactions will merge into each other at an energy of 1 TeV, which is unrealistic for LBL experiments. Such energetic neutrinos may be detectable in Antares \cite{Asl}, Nestor \cite{Resv}, Amanda \cite{And}, and Baikal \cite{Zha}.
 In Figs.~\ref{taucon} we show $\nu_\tau$ induced cross sections for charged and neutral currents. The energy scale is now expanded to show clearly the threshold effect. Our results agree well with those in Ref. \cite{Hall}. We notice that the slope of the neutral current reactions remains constant also for low energies, while the production of $\tau$-leptons begins between $4-5$ GeV and their strength reaches large values above 8 GeV. The high energy values of charged and neutral current reactions are comparable. Figs.~\ref{taunuc} and ~\ref{ncnuc} show the slope of cross sections on iron target. We included in this case the nuclear corrections which turn out to be small (of order $5-7\%$). The main characteristic is the threshold dependence of the charged current reactions. Thus if the experiments can measure neutral current reactions for low energies, $E_\nu < 5$ GeV, they should observe a linear energy dependence of the events coming from $\nu_\mu$ and $\nu_\tau$ neutrinos, because even after the oscillation $\nu_\mu\rightarrow\nu_\tau$ their contributions are equal. If, on the other hand, the oscillation is $\nu_\mu\rightarrow\nu_s$ then there should be a decrease of the cross section in the far away detector, since the sterile neutrinos do not contribute. This decrease should be a function of $E_\nu$ and should be maximal when $\frac{\delta m^2 L}{4 E_\nu}\approx \frac{\pi}{2}$. For comparison with other reactions we plot in Fig.~\ref{tot}, the slope of the CC cross sections for the sum of the three types of reactions. We note that at high energies DIS dominates. The QE and RES give a noticeable contribution around 5 GeV where a kink in the slope is visible.    

\subsubsection {quasi-elastic scattering}

\begin{figure}[b]
\noindent
\begin{minipage}[b]{.49\linewidth}
\centering\epsfig{figure=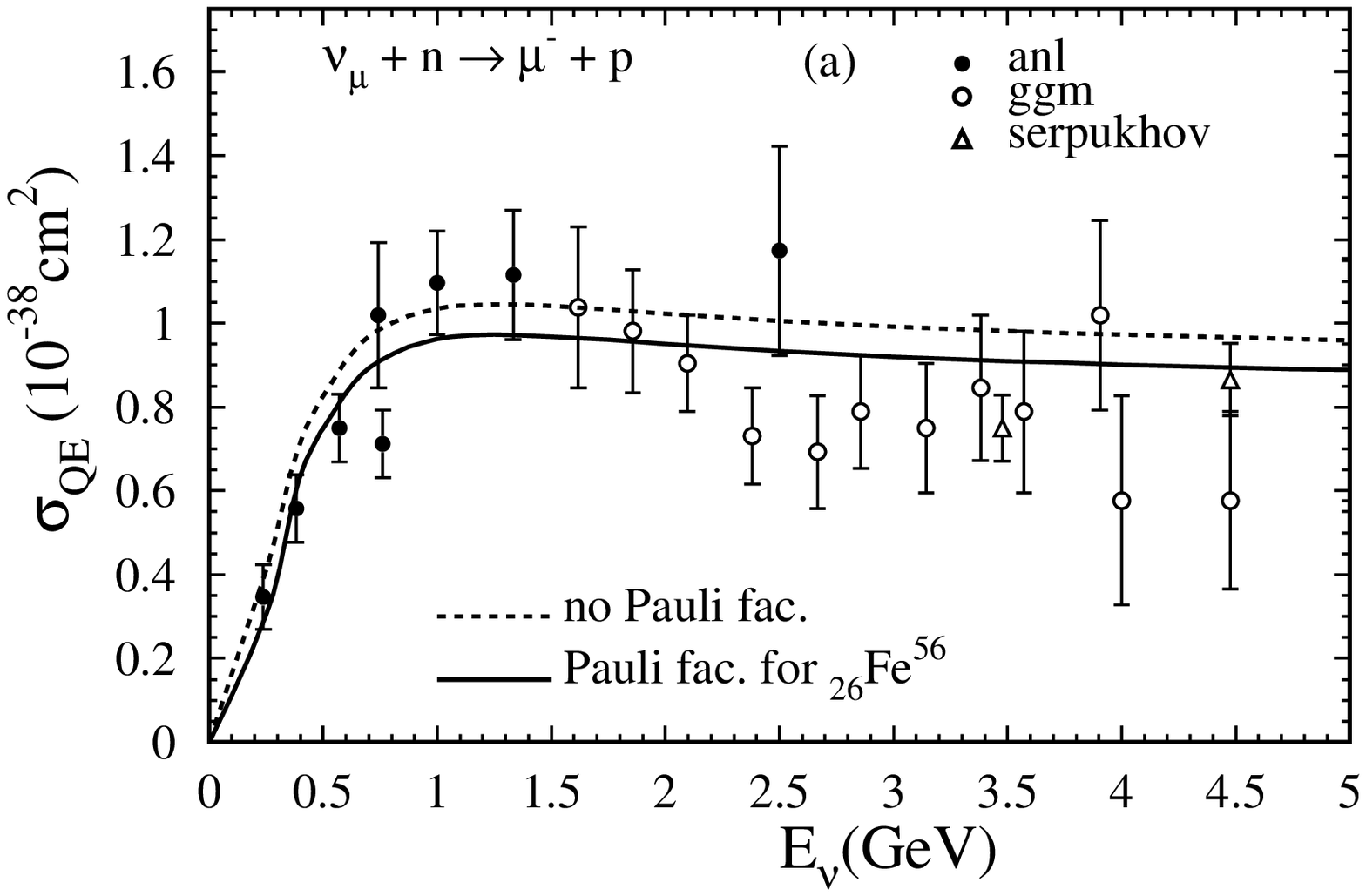,height=3.0in,width=\linewidth}
\end{minipage}
\begin{minipage}[b]{.49\linewidth}
\centering\epsfig{figure=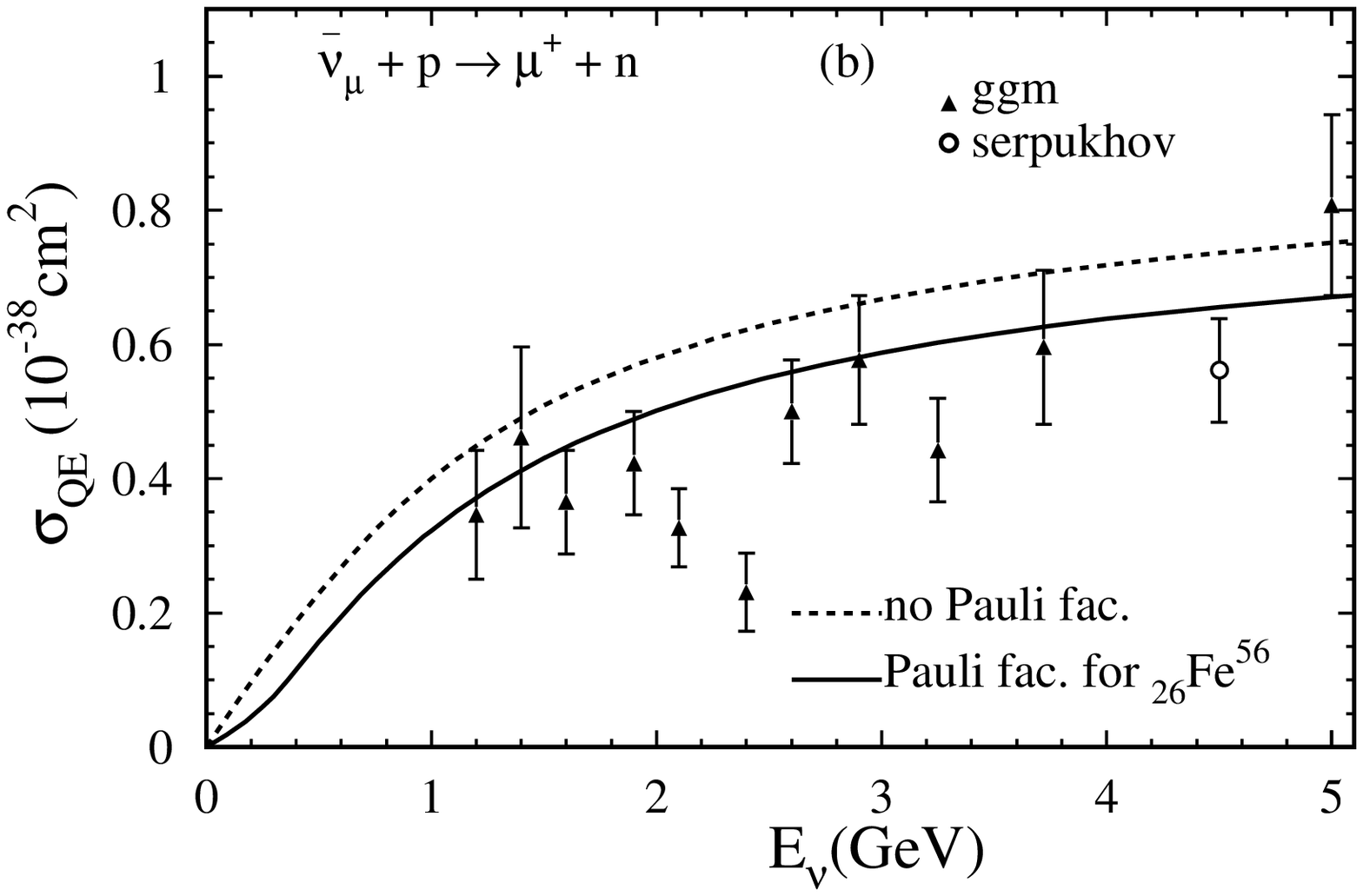,height=3.0in,width=\linewidth}
\end{minipage}
\vspace{-1.0cm}
\caption{The cross section of QE for the (a) $\nu_\mu + n \rightarrow \mu^- + p $ and (b)  $\bar{\nu}_\mu + p \rightarrow \mu^+ + n $ process plotted versus the incoming neutrino energy with and without Pauli factor. The data points are from ANL \protect\cite{Bari}, GGM \protect\cite{Bonetti} and Serpukhov \protect\cite{Belikov}.} \label{qemu}
\end{figure}

In Figs.~\ref{qemu}\,\,-~\ref{qenc} we present our results for QE scattering. Figs.~\ref{qemu} show cross section on free protons induced by $\nu_\mu$ and $\bar{\nu}_\mu$'s. The cross sections reach a constant asymptotic value at an energy of 2 GeV. We included also the Pauli factor whose effect is small. The data are closer to the curves which include the Pauli factor. In Figs.~\ref{qetau} we show the charged current cross section induced by $\nu_\tau$'s. The threshold dependence is again prominent and the cross sections have an energy dependence even at $E_{\nu_\tau} \approx 10$ GeV. The Pauli factor effect is in this case small, bringing a decrease about 10$\%$. Finally, the neutral current cross sections rise quickly to their asymptotic values (Figs.~\ref{qenc}) which are approximately 10$\%$ of the $\nu_\mu$ CC cross section. Thus the threshold effects can distinguish between $\nu_\mu$ and $\nu_\tau$ interactions. The neutral current events should not show a threshold effect if the oscillation is $\nu_\mu\rightarrow\nu_\tau$ and there should be no $\tau$-leptons if the oscillation is $\nu_\mu\rightarrow \nu_s$. 

\subsubsection{Resonance Production}

\begin{figure}
\begin{minipage}[b]{.49\linewidth}
\centering\epsfig{file= 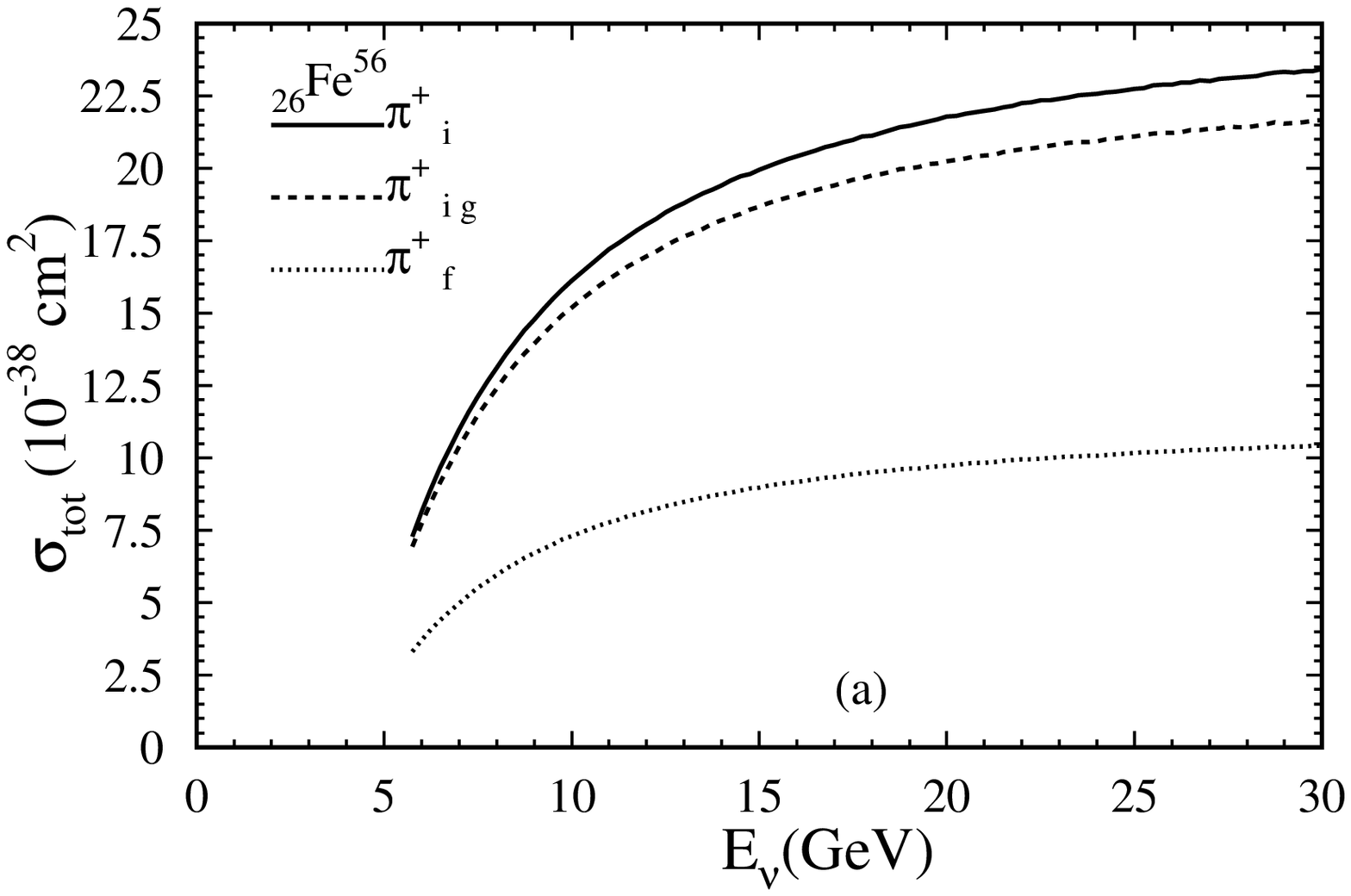,height=3.6in,width=\linewidth}
\end{minipage}
\begin{minipage}[b]{.49\linewidth}
\centering\epsfig{file= 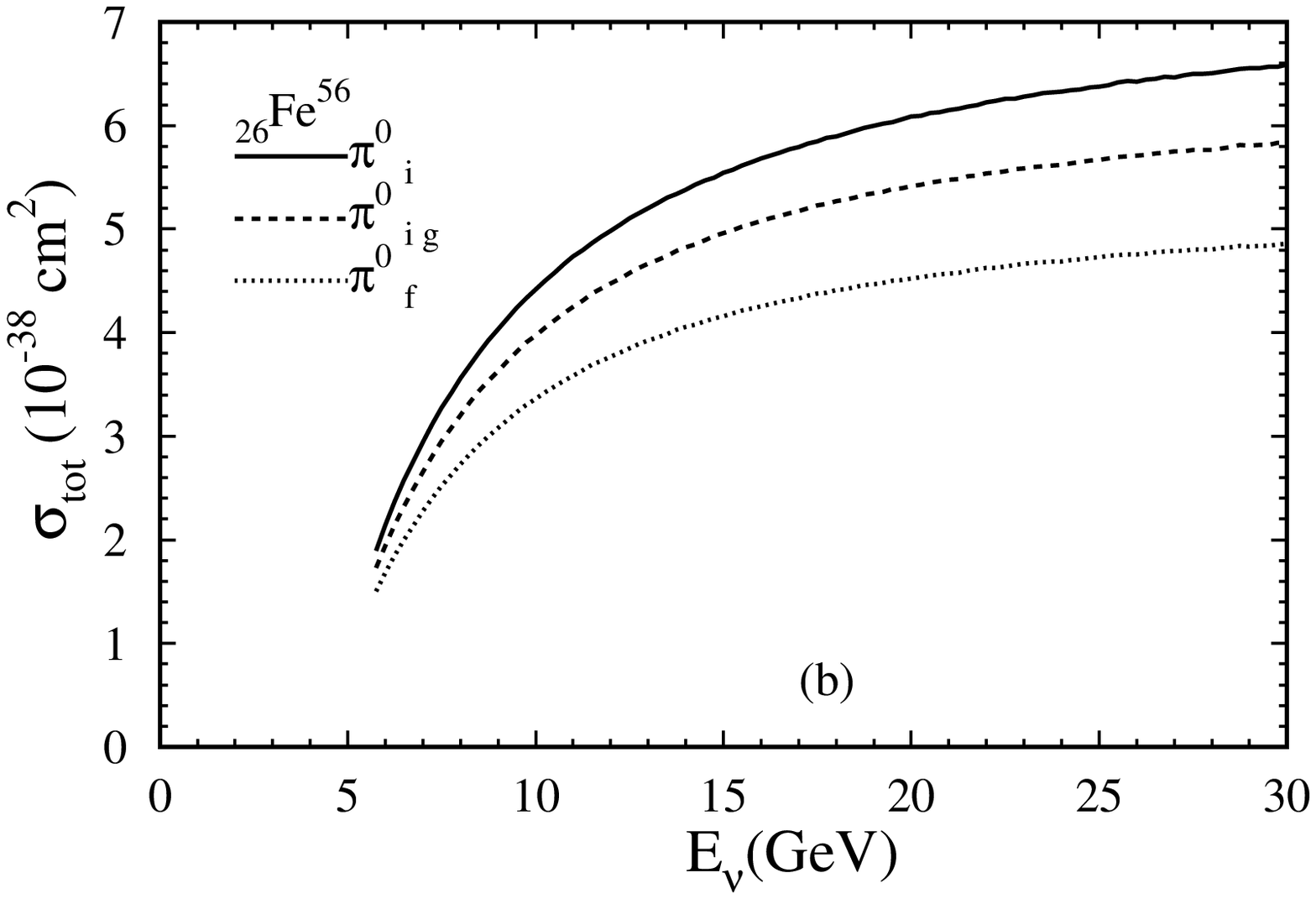,height=3.6in,width=\linewidth}
\end{minipage}
\caption{The cross section of RES for (a) positively charged  and (b) neutral pions produced on iron targets. The solid, dashed and dotted lines represent respectively the cross sections without any nuclear correction, including only the Pauli production factor $g$ and including all nuclear corrections.}\label{res1}
\end{figure}

Resonance production induced by $\nu_\mu$ neutrinos was studied in our previous article \cite{Yu}. Here we extend the study to the production of $\tau$'s and their associated threshold effects. Resonance production provides an additional signature, because in this case there is also a pion in the final state. For CC reactions the signal will be a lepton and a charged pion in the final state, while the neutral currents will search for a single pion. The nuclear effects are expected to be larger because the pions have a chance of rescattering within the nucleus \cite{Adler}. Fig.~\ref{res} (a) shows the various channels produced in the reactions
\begin{eqnarray}
\nu_\tau + N \rightarrow \tau^- + N +\pi^{+,0}
\end{eqnarray}
and the corresponding reactions with anti-neutrinos are shown in Fig.~\ref{res} (b). The threshold is again at $E_\nu \approx 5.6$ GeV. The cross sections grow rapidly now, reaching their asymptotic values at $E_\nu\approx 15$ GeV. For a heavy target like iron the same features appear, but now there are nuclear effects like rescattering and charge-exchange, bringing a substantial reduction: a factor of 2 for $\pi^+$ (Fig.~\ref{res1} (a)) and a factor of 30 $\%$ for $\pi^0$ (Fig.~\ref{res1} (b)). Similarly there is reduction by a factor $\sim$ 2 for $\pi^-$ in the reaction 
\begin{eqnarray}
\bar{\nu}_\tau + N\rightarrow \tau^+ +N +\pi^- 
\end{eqnarray}
shown in Fig.~\ref{resan}. In Figs.~\ref{opp}\,\,-~\ref{fem} we show the differential cross sections $d\sigma/dE_\pi$ for the reactions $\nu_\tau(\bar{\nu}_\tau) + N\rightarrow \tau^-(\tau^+) +N +\pi^{\pm,0}$ on three nuclear targets, $_8{\rm O}^{16}, _{18}{\rm Ar}^{40}$ and $_{26}{\rm Fe}^{56}$. The pion spectra are significantly reduced reflecting the rescattering of the pions. In some cases the reduction is a whole factor of 2. It is also interesting to note that the reaction of the various channels is different than in the case of neutral currents \cite{Yu}.

\section{Event Rates}\label{sec:four}

We calculated the $\tau$-lepton event rates of the total cross section (RES+QE+DIS), concentrating on the OPERA LBL experiment \cite{Opera}. The number of  observable $\nu_\tau$ charged current events, $N_\tau$, using the CERN-NGS neutrino beam is given by the following equation:
\begin{equation}
N_\tau =  A \int \,\phi_{\nu_\mu}(E_\nu) P_{osc}(\nu_\mu\rightarrow\nu_\tau) \sigma^{CC}_{\nu_\tau} (E_\nu) Br(\tau\rightarrow  {\rm lepton(l) ,hadron(h)}) \epsilon(E_\nu) dE_\nu, 
\end{equation} 
where  $\phi_{\nu_\mu}$ is muon neutrino flux at the Gran Sasso detector which  we took from \cite{NGS} and $ \sigma^{CC}_{\nu_\tau}$ is the charged current total cross section for the $\nu_\tau $ from our theoretical calculation. For the neutral current total cross section we replace $\sigma^{CC}_{\nu_\tau}$ to  $\sigma^{NC}_{\nu}$. The neutrino flux is appropriately normalized so that $A$ is the total number of active protons plus neutrons in the target. It is referred to as the active target mass $A$ and is given by $A = N_A \times 10^9 \times M_d\times N_p\times N_y$, where $N_A$ is Avogadro's number. We take the detector mass $M_d = 1$ kton, the number of years for data taking $N_y$ = 4 and the number of protons on target per year $N_p = 4.5\times 10^{19}$ pot/year.   
In the two flavor mixing scheme we took the probability of $\nu_\mu \rightarrow \nu_\tau$ given by the following equation:
\begin{equation}
P_{osc}(\nu_\mu\rightarrow\nu_\tau)  = \sin^2 2 \theta\sin^2\Big(\frac{1.27 \delta m^2 L}{E_\nu}\Big),
\end{equation} 
with $\sin^2 2 \theta = 1$ and the distance $L$ from CERN to Gran Sasso Laboratory is 730 km. We considered the neutrino energy range as 1 GeV $\leq E_\nu\leq$ 30 GeV and took $\delta m^2 = 10^{-3} - 10^{-2}\,{\rm eV}^2$. We adopted the branching ratios $Br(\tau\rightarrow {\rm lepton(l) ,hadron(h)})$ and the detector efficiency $\epsilon$ of the $\nu_\tau$ events from Ref.\cite{Gago}. We did not consider background because the number of such events is expected to be very small, as can be seen from \cite{Opera}.

\subsection{results}

We present the number of events, $N_\tau$, for charged currents (Tables~\ref{t1} and~\ref{t2}) and for neutral currents (Tables~\ref{t3} and~\ref{t4}). In order to distinguish the various channels we give events for DIS, separately, and also the sum of DIS+QE+RES. Table~\ref{t1} shows the number of charged current events for the oscillation parameters at two confidence levels with $\delta m^2 \simeq 10^{-3}-5\cdot 10^{-3}\,{\rm eV}^2$ and $\sin^2 2\theta\geq 0.89$ taken from a recent global analysis \cite{Gon}. In the first column is the total number of events without nuclear corrections. The second column shows events for DIS with nuclear corrections and the third one the total number of events with nuclear corrections. 
Table~\ref{t2} shows charged currents events as a function of $\delta m^2$. The various columns include events classified in the same way as in Table~\ref{t1}.\\
\indent In Tables~\ref{t3} and~\ref{t4} we present the neutral currents events. Table~\ref{t3} includes the number of events again for two confidence levels and the various columns have the same meaning as above. Finally, Table~\ref{t4} shows the number of events as a function of $\delta m^2$. 
The event numbers for charged current channels in Ref. \cite{Gago,Rubbia} are in reasonable agreement with our results. For the total number of events in QE+RES+DIS the reduction from nuclear corrections is $10-15\%$, mainly because nuclear corrections are significant for the resonance channels. Nuclear effects reduce RES and QE processes, but remain negligible for DIS processes. The contribution of QE and RES production to $\tau$-appearance events, in charged currents, is approximately $20 - 24\%$ and, in neutral currents, about $13-15\%$.  

\section{Conclusions}\label{sec:five}

\begin{figure}
\centerline{\psfig{figure=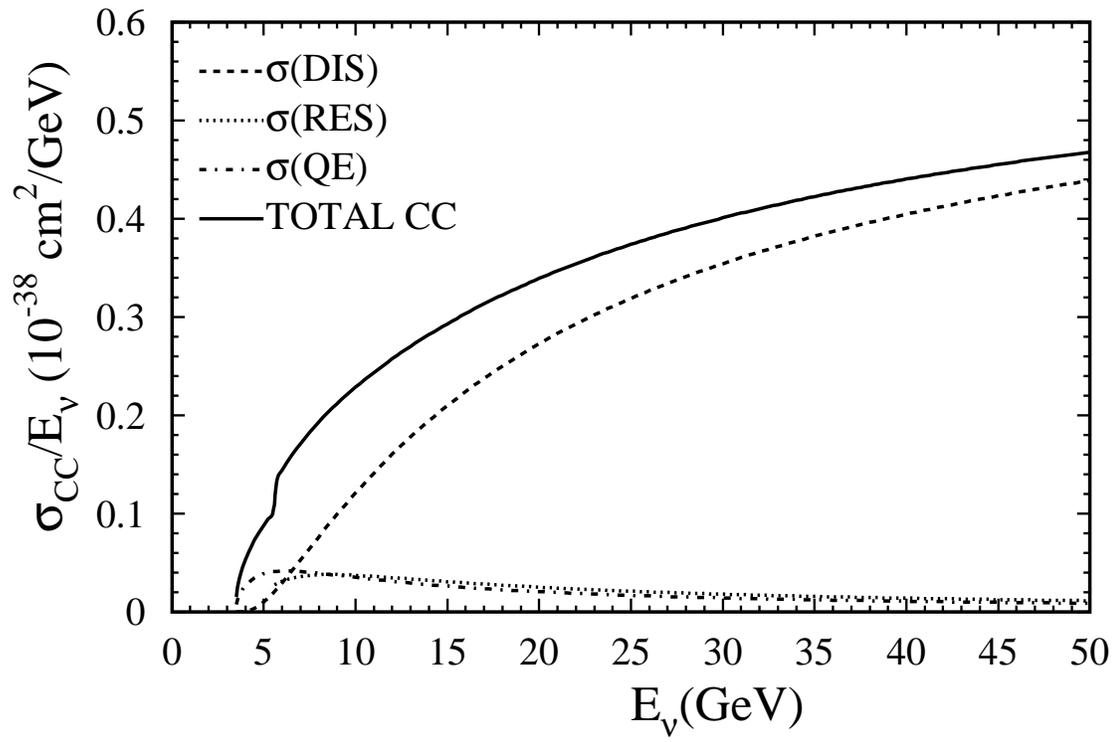,height=5.5in,angle=0}}
\caption{All the types of $\nu_\tau$ charged current cross sections for an isoscalar target plotted as a function of neutrino energy with the normalization of $1/E_\nu$.}\label{tot}
\end{figure}

\begin{figure}[b]
\begin{minipage}[b]{.49\linewidth}
\centering{\epsfig{figure=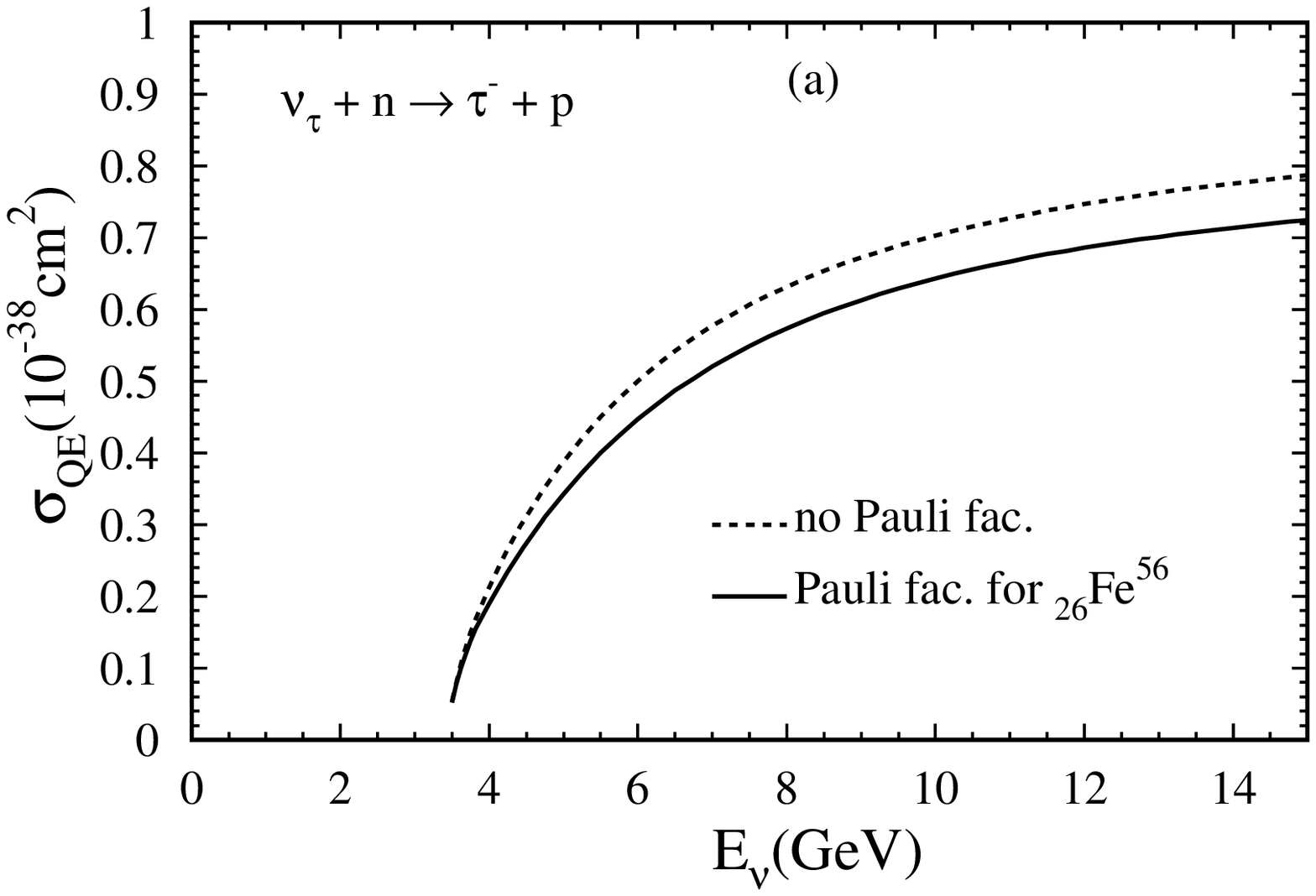,width=\linewidth}}
\end{minipage}
\begin{minipage}[b]{.49\linewidth}
\centering{\epsfig{file=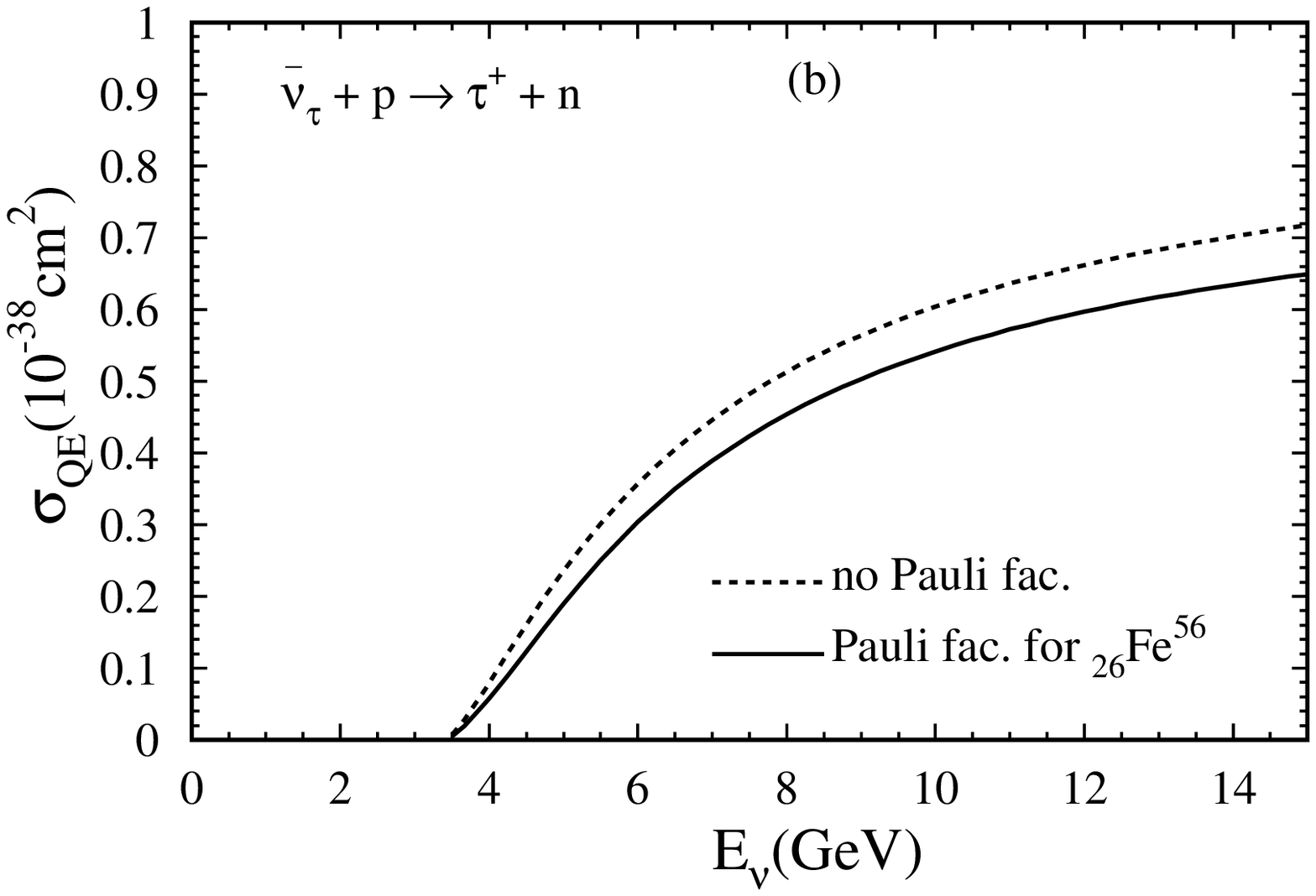,width=\linewidth}}
\end{minipage}
\vspace{-1.0cm}
\caption{The cross section of QE for the (a) $\nu_\tau + n \rightarrow \tau^- + p$ and (b) $\bar{\nu}_\tau + p \rightarrow \tau^+ + n$ process plotted versus the incoming neutrino energy with and without Pauli factor.}\label{qetau}
\end{figure}

\begin{figure}[b]
\begin{minipage}[b]{.49\linewidth}
\centering\epsfig{figure=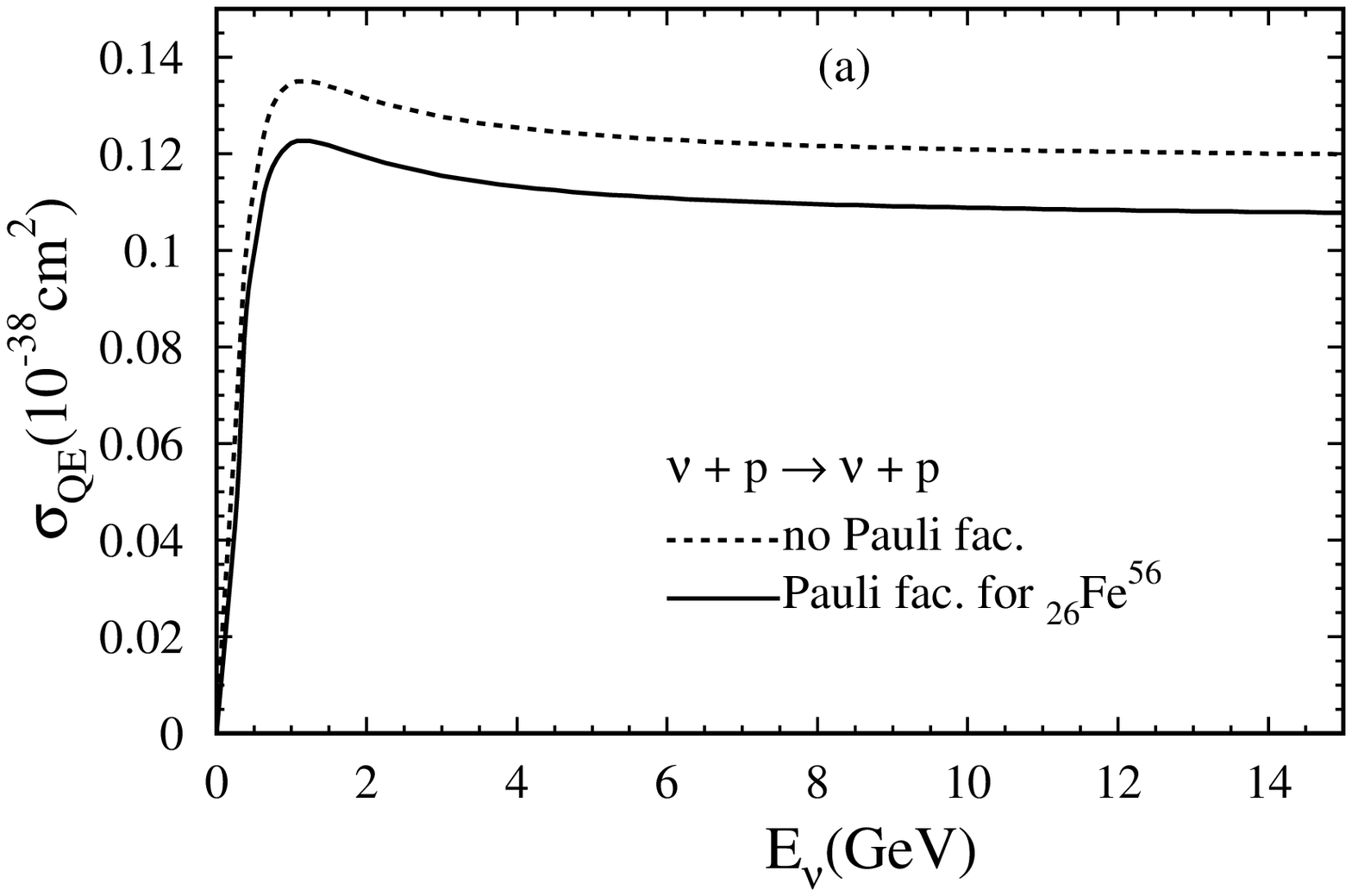,width=\linewidth}
\end{minipage}
\begin{minipage}[b]{.49\linewidth}
\centering\epsfig{file=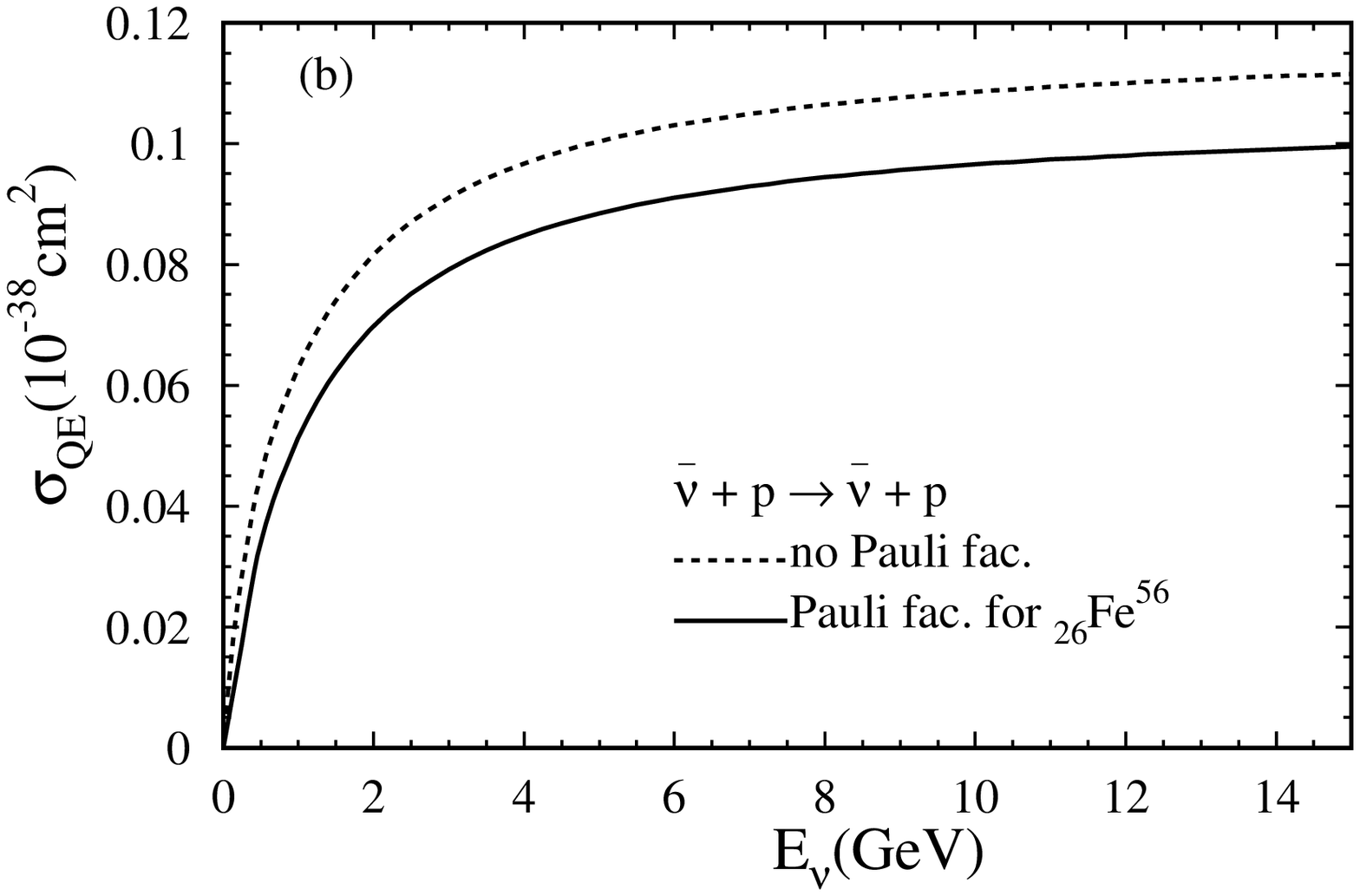,width=\linewidth}
\end{minipage}
\vspace{-1.0cm}
\caption{The cross section of QE for the (a) $\nu + p \rightarrow \nu + p$ and (b) $\bar{\nu}_\tau + p \rightarrow \bar{\nu} + p$ processes plotted versus the incoming neutrino energy with and without Pauli factor.}\label{qenc}
\end{figure}

\begin{figure}
\begin{minipage}[b]{.49\linewidth}
\centering\epsfig{file=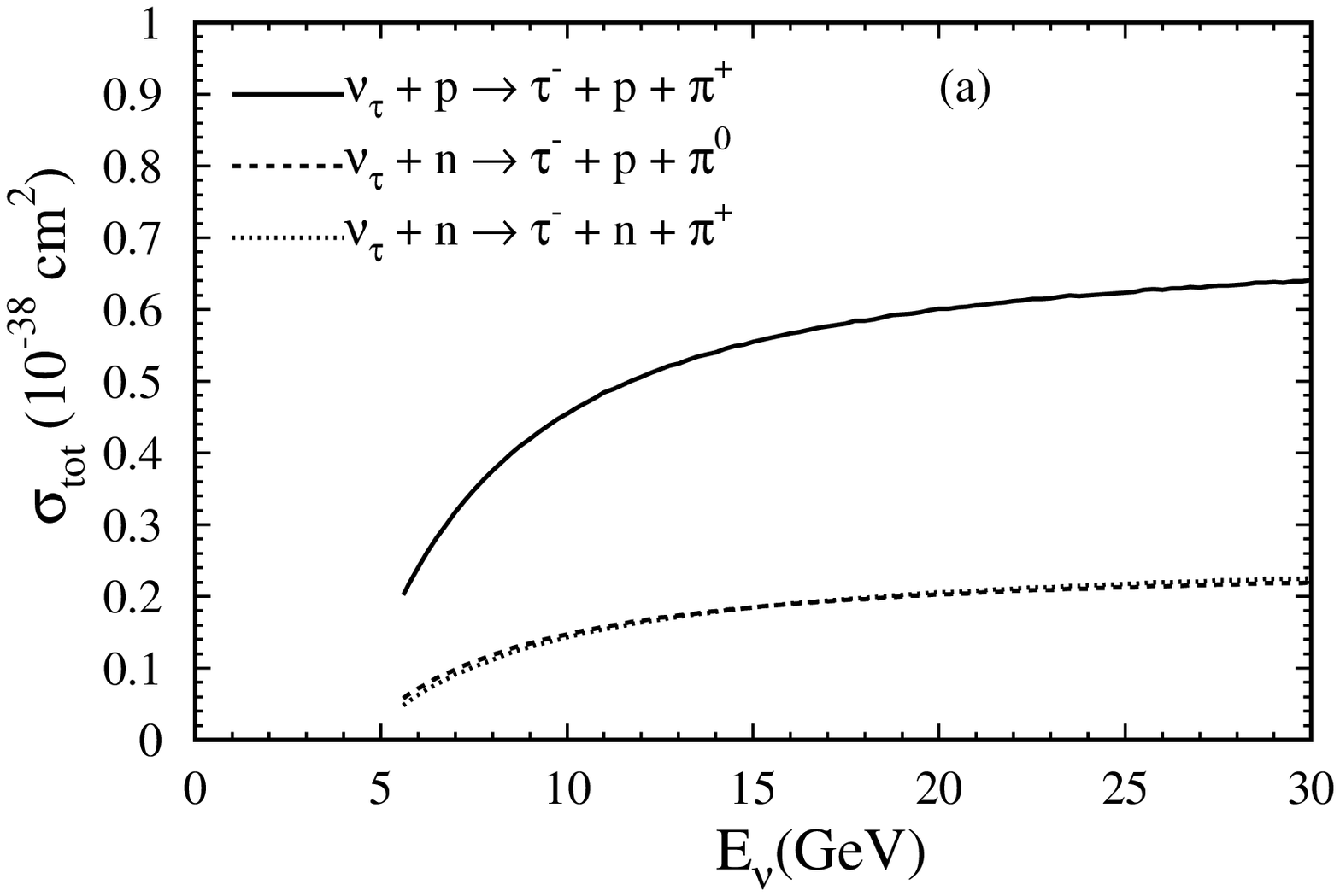,height=3.6in,width=\linewidth}
\end{minipage}
\begin{minipage}[b]{.49\linewidth}
\centering\epsfig{file=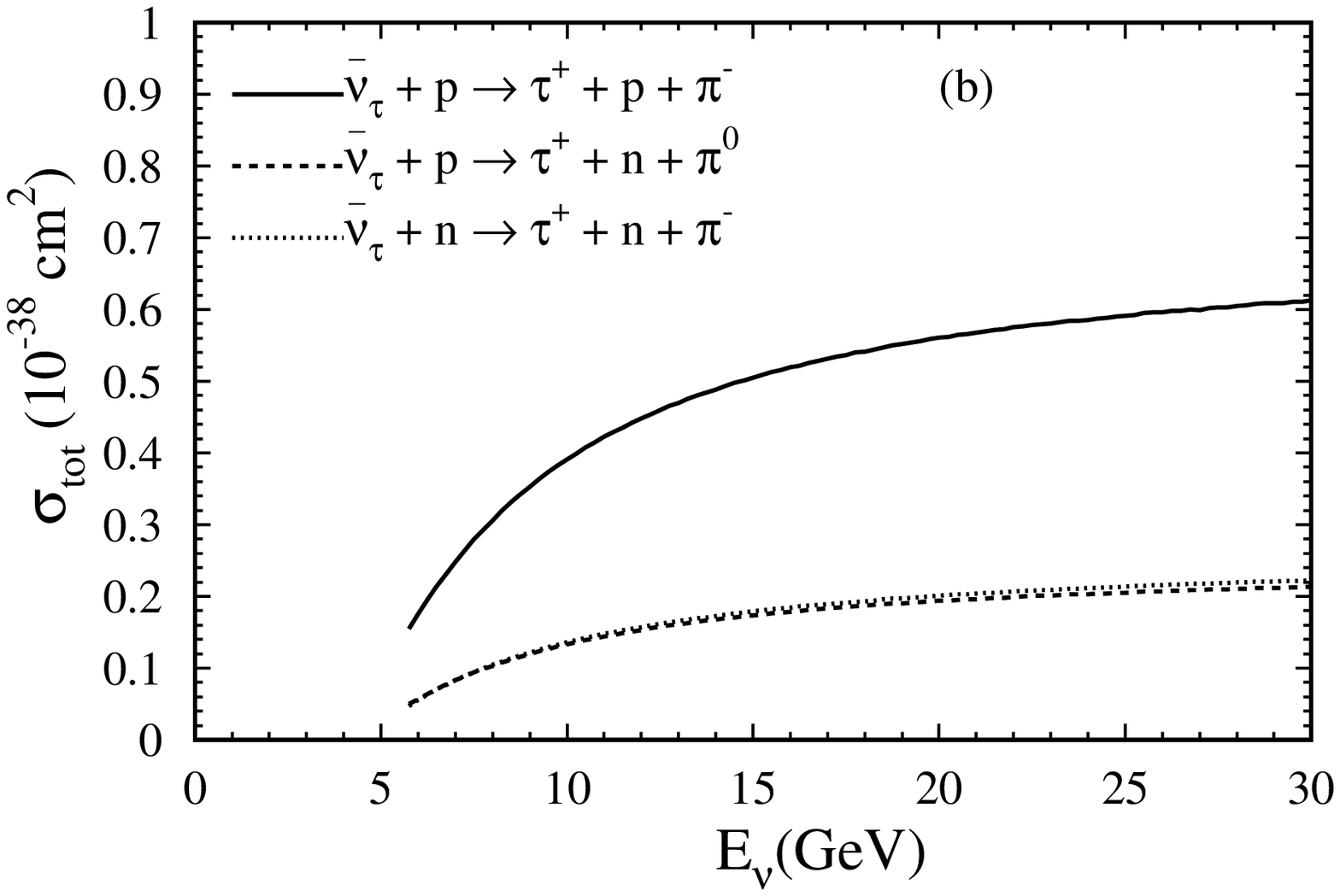,height=3.6in,width=\linewidth}
\end{minipage}
\vspace{-1.0cm}
\caption{The cross section of RES for (a) $\nu_\tau +N\rightarrow \tau^- + N +\pi^{+,0 }$ and (b) $\bar{\nu}_\tau +N\rightarrow \tau^+ + N +\pi^{-,0 }$ plotted versus the incoming neutrino energy.}\label{res}
\end{figure}

\begin{figure}
\centerline{\psfig{figure=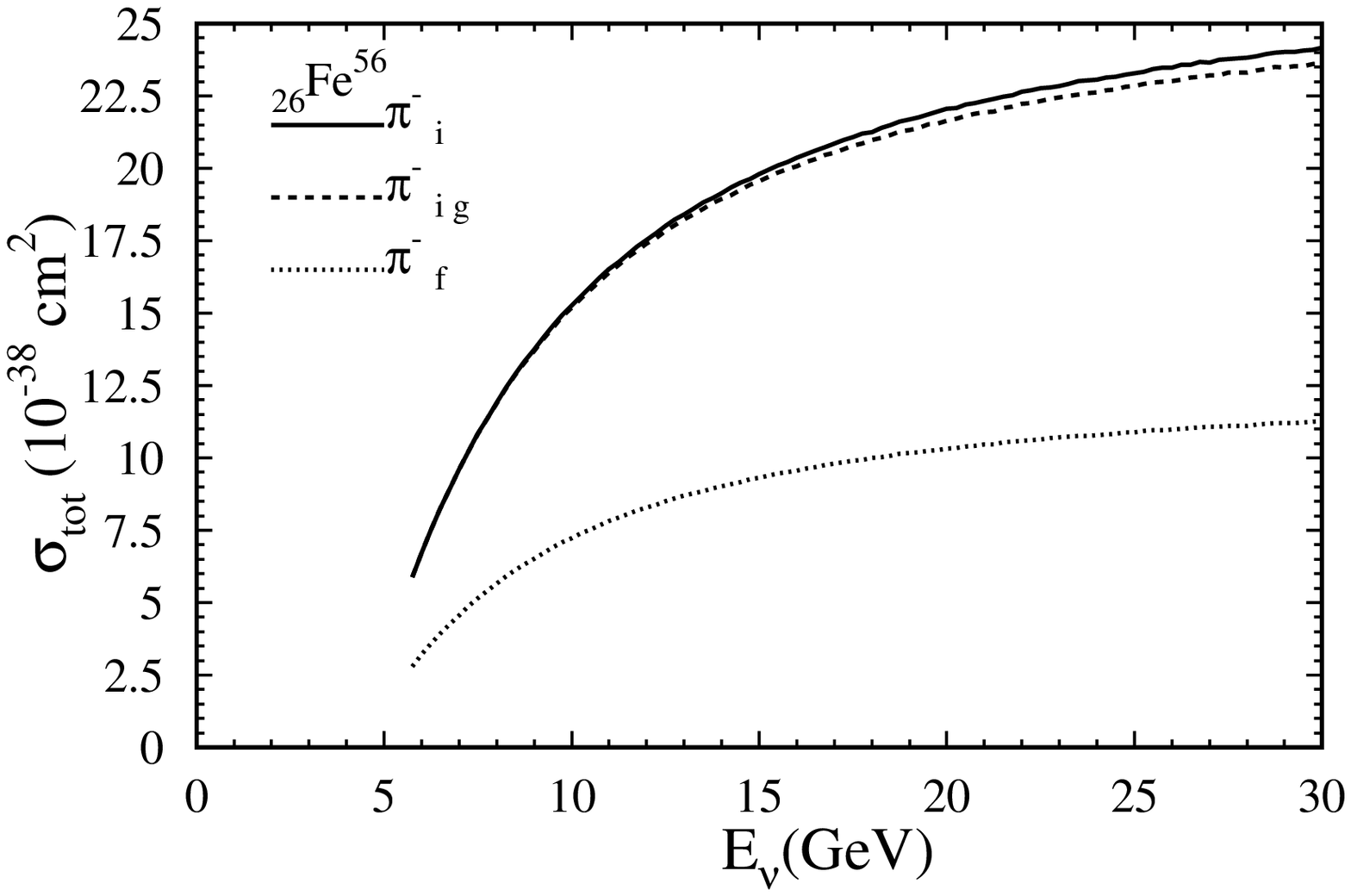,height=3.6in,angle=0}}
\caption{The cross section of RES for negatively charged  and neutral current pions produced on iron targets. The solid, dashed and dotted lines represent respectively the cross sections without any nuclear correction, including only the Pauli production factor $g$ and including all nuclear corrections.}\label{resan}
\end{figure}

\begin{figure}
\begin{minipage}[b]{.49\linewidth}
\centering\epsfig{file=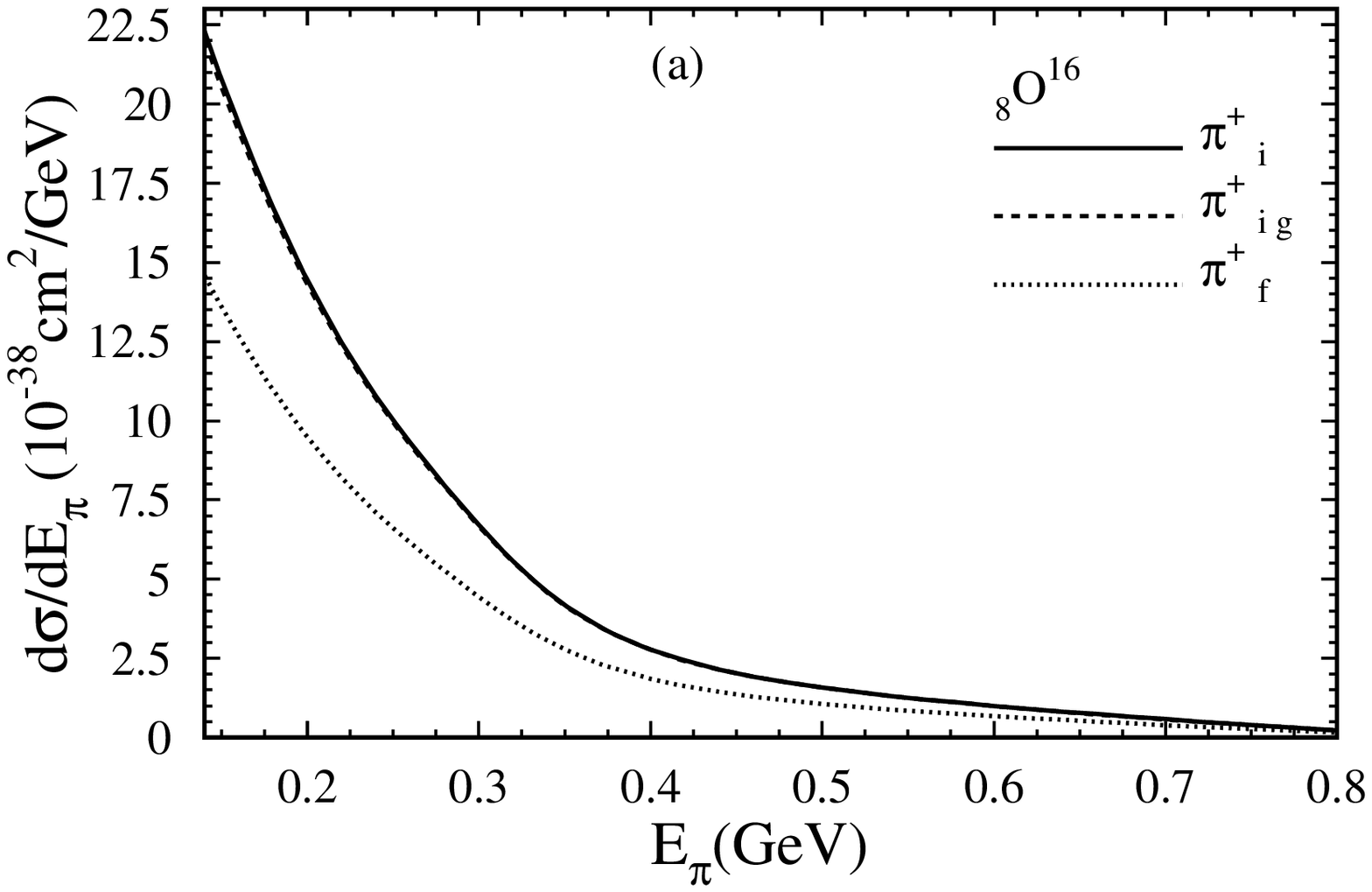,height=3.6in,width=\linewidth}
\end{minipage}
\begin{minipage}[b]{.49\linewidth}
\centering\epsfig{file=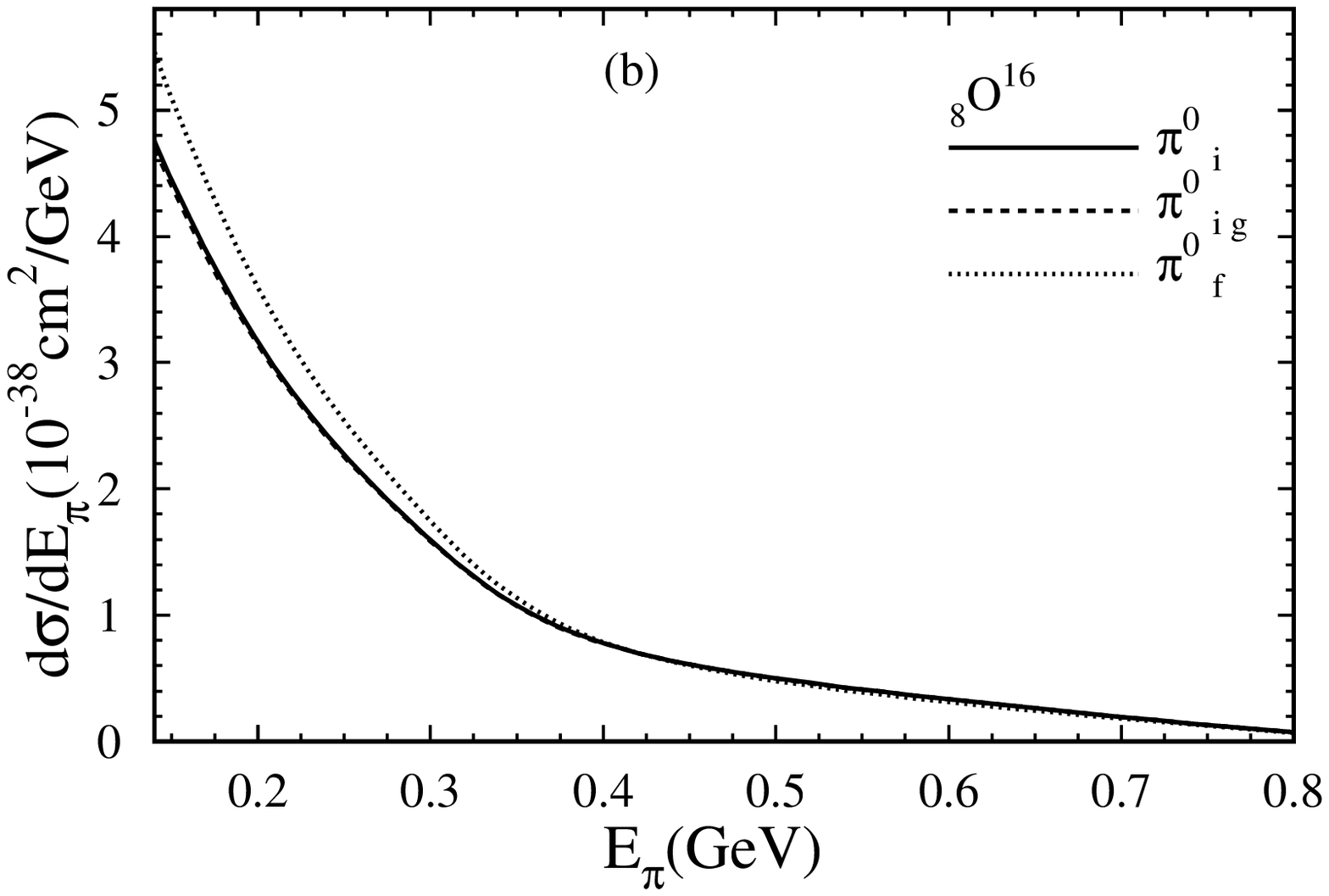,height=3.6in,width=\linewidth}
\end{minipage}
\vspace{-1.0cm}
\caption{Pion energy distribution for (a) positively charged pion and (b) neutral pion produced on oxygen targets. The solid, dashed and dotted lines represent respectively the pion energy distribution without any nuclear correction, including only the Pauli production factor $g$ and including all nuclear corrections.}\label{opp}
\end{figure}

\begin{figure}
\centerline{\psfig{figure=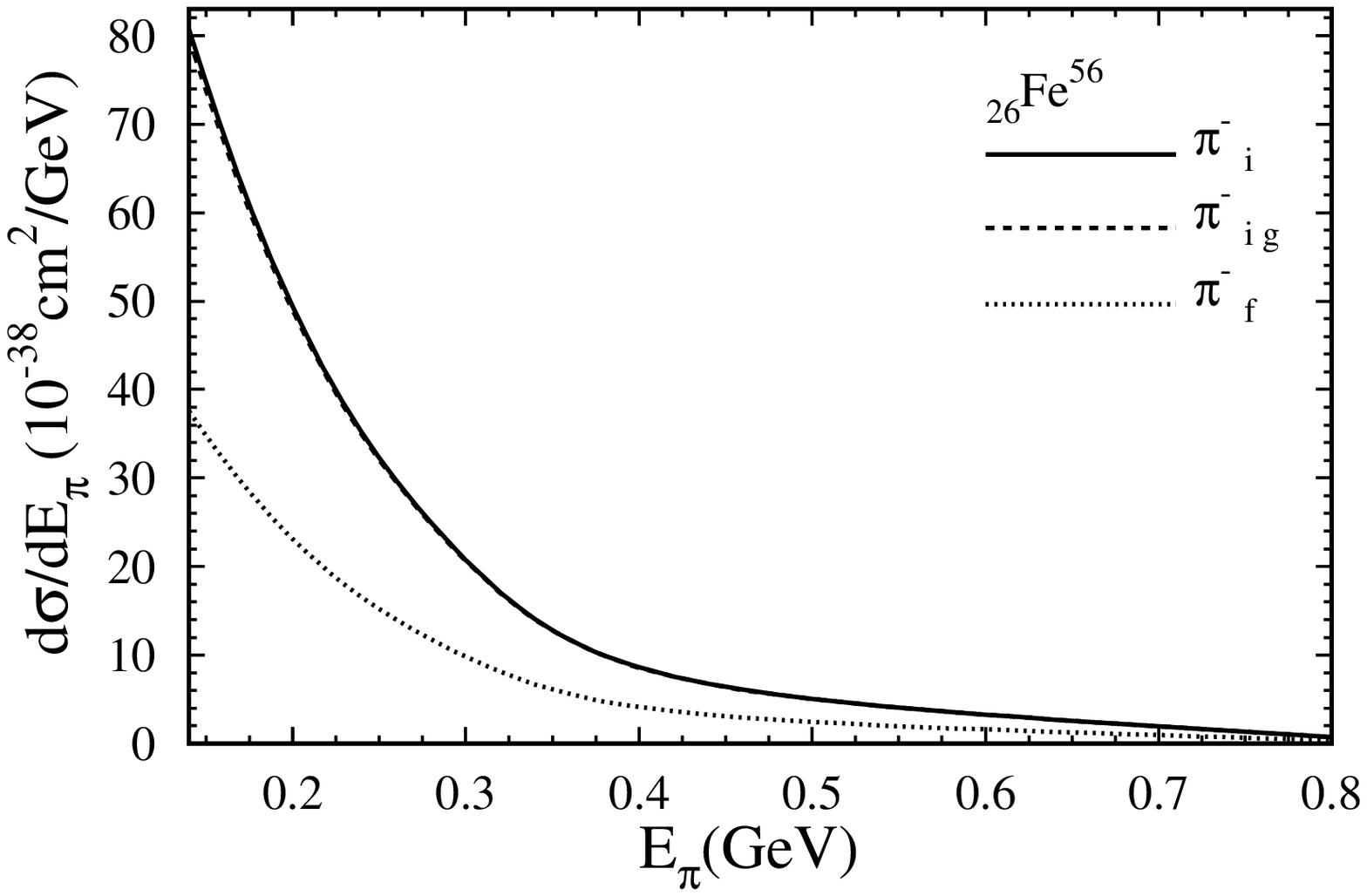,height=3.6in,angle=0}}
\caption{Pion energy distribution for negatively charged pions produced on iron targets. The solid, dashed and dotted lines represent respectively the pion energy distribution without any nuclear correction, including only the Pauli production factor $g$ and including all nuclear corrections.}\label{fem}
\end{figure}

Neutrino oscillation experiments face the problem that the number of events is very small. This limitation is more severe for $\tau$-appearance experiments, which motivated to design experiments with heavy nuclei as targets. The number of events will now increase substantially, given by $(A-Z)\sigma_n + Z \sigma_p$ with $\sigma_{p,n}$ being the cross sections on free protons and neutrons, respectively. This substantial increase is slightly complicated by nuclear target effects. In this paper we calculated several cross sections and showed how the nuclear effects can be understood and compensated for. \\
\indent The role of the various reactions is distinct. For energies $E_\nu < 2.5$ GeV the $\mu^-$ - production and neutral current reactions receive comparable contributions from three types of reactions: quasi-elastic, resonance production and deep inelastic scattering. The analysis of the data must include all three of them and try to identify unique signatures.\\
(1) In quasi-elastic scattering there is a single nucleon in the final state, which is unique but hard to detect.\\
(2) In resonance production there is a nucleon and a pion, whose decay gives a unique signature. The produced pions may be further identified and confirmed by their specific energy spectra as plotted in Figs.~\ref{opp}\,\,-~\ref{fem}.\\
(3) For $E_\nu > 2.5$ GeV deep inelastic scattering dominates the $\nu_\mu$ reactions. For comparison, the $\tau$-lepton events have a characteristic threshold dependence. There are no $\tau$-leptons produced for $E_\nu < 5.6$ GeV. For the $\nu_\tau$ beams quasi-elastic and resonance production are important for energies $E_\nu$ up to 6.5 GeV to 7.0 GeV. Above this energy (see Fig.~\ref{tot}) the deep inelastic reaction dominates. \\
\indent In the resonance region our results of the nuclear effects for neutral currents were presented in Ref. \cite{Yu}. In this article we extended the calculations to $\tau$-appearance experiments and we can summarize them as follows.\\
(1) Nuclear effects are very small for deep inelastic reactions and can be neglected.\\
(2) For quasi-elastic scattering the main effect is the Pauli suppression factor, which reduces the rates by 10-12 $\%$ (See Figs.~\ref{qetau} and \ref{qenc}).\\
(3) Nuclear corrections are substantial in single-pion production at the resonance region. They vary from channel to channel and for this reason we produced Figs.~\ref{res}\,-\ref{resan} showing the production cross sections and Figs.~\ref{opp}\,-\ref{fem} showing the pion spectra. A striking feature in all of the cross sections is the $\tau$-lepton threshold.\\ 
\indent We also made an extensive search of earlier publications trying to find data for possible experimental comparisons. Inspite of our efforts we could not find data for a meaningful test of nuclear corrections. Thus it is advisable for the nearby detectors of the LBL experiments to collect data on heavy nuclei and test the models \cite{Adler,Oset}. The required comparisons are evident from the present and previous articles \cite{Yu,Adler}. 
            
\acknowledgements
In the progress of this work we profited from the expertise of our colleagues. One of us (EAP) wishes to thank S.L.Adler at the Institute for Advanced Study for the hospitality, where part of the work was performed. Another of us (JYY) thanks L.P.Singh for helpful discussions. The work of JYY is supported by the German Federal Ministry of Science (BMBF) under contract 05HT9PEA5.   
\pagebreak



\begin{table}
\caption{The number of events $N_\tau$ at OPERA for the cross section for DIS and for the total cross section (QE+RES+DIS) with and without nuclear corrections. The table is for charged current channels with the $90\%$ and $99\%$ C.L. parameter set at the value $\delta m^2 \simeq 10^{-3}-5\cdot 10^{-3}\,{\rm eV}^2$  and  $\sin^2 2\theta\geq 0.89$ of~[\ref{Gon}].}
\begin{tabular}{cc|cccc} 
 & &  $N_\tau$  & $N_\tau(nucl. for DIS)$ & $N_\tau(nucl. for DIS+RES+QE)$\\ \hline
$90\% C.L.$ & $min$ &  4.06   &  2.72     &  3.56\\ 
            & $max$ & 30.72   &  20.92    & 26.67\\ \hline
$99\% C.L.$ & $min$ &  2.33   &  1.55     &  2.04  \\ 
            & $max$ & 41.13   & 28.66     & 35.71\\
\end{tabular} \label{t1}
\end{table}

\begin{table}
\caption{The number of events $N_\tau$ at OPERA for the cross section for DIS and for the total cross section (QE+RES+DIS) with and without nuclear corrections. These are charged current channels with $\sin^2 2 \theta = 1$ and various $\delta m^2$.}
\begin{tabular}{c|cccc}
$\delta m^2 (eV^2)$ & $N_\tau$  & $N_\tau(nucl. for DIS)$& $N_\tau(nucl. for DIS+RES+QE)$\\ \hline
1.5 $\times 10^{-3}$ &  2.69   &  1.8   &  2.36\\
3.0 $\times 10^{-3}$ & 11.34   &  7.62  &  9.87\\
3.5 $\times 10^{-3}$ & 15.57   & 10.49  & 13.53\\
4.5 $\times 10^{-3}$ & 25.80    & 17.51 & 22.39\\
5.0 $\times 10^{-3}$ & 31.72   & 21.60   & 27.53\\
\end{tabular}\label{t2}
\end{table}

\begin{table}
\caption{The number of events $N_\tau$ at OPERA for the cross section for DIS and for the total cross section (QE+RES+DIS) with and without nuclear corrections. These are neutral current channels with the $90\%$ and $99\%$ C.L parameter set at the value $\delta m^2 \simeq 10^{-3}-5\cdot 10^{-3}\,{\rm eV}^2$ and $ \sin^2 2\theta\geq 0.89$ of~[\ref{Gon}].}
\begin{tabular}{cc|cccc} 
 & & $N_\tau$  & $N_\tau(nucl. for DIS)$ & $N_\tau(nucl. for DIS+RES+QE)$\\ \hline
$90\% C.L.$ & $min$ &  6.89  &  5.26     &  6.15 \\ 
            & $max$ & 42.34  & 33.56     &  37.91\\ \hline
$99\% C.L.$ & $min$ &  4.10  &  3.11     &  3.66 \\ 
            & $max$ & 55.21  & 43.98     &  49.46\\
\end{tabular}\label{t3}
\end{table}

\begin{table}
\caption{The number of events $N_\tau$ at OPERA for the cross section for DIS and for the total cross section (QE+RES+DIS) with and without nuclear corrections. These are neutral current channels with $\sin^2 2 \theta = 1$ and various $\delta m^2$.}
\begin{tabular}{c|cccc}
$\delta m^2 (eV^2)$ & $N_\tau$ & $N_\tau(nucl. for DIS)$ & $N_\tau(nucl. for DIS+RES+QE)$\\  \hline
1.5 $\times 10^{-3}$ &  4.72    &   3.58    &  4.21 \\
3.0 $\times 10^{-3}$ & 17.45    &  13.54    &  15.59\\
3.5 $\times 10^{-3}$ & 23.11    &  18.05    &  20.66\\
4.5 $\times 10^{-3}$ & 36.26    &  28.63    &  32.45\\
5.0 $\times 10^{-3}$ & 43.66    &  34.62    &  39.1 \\
\end{tabular}\label{t4}
\end{table}


\begin{figure}
\centerline{\psfig{figure=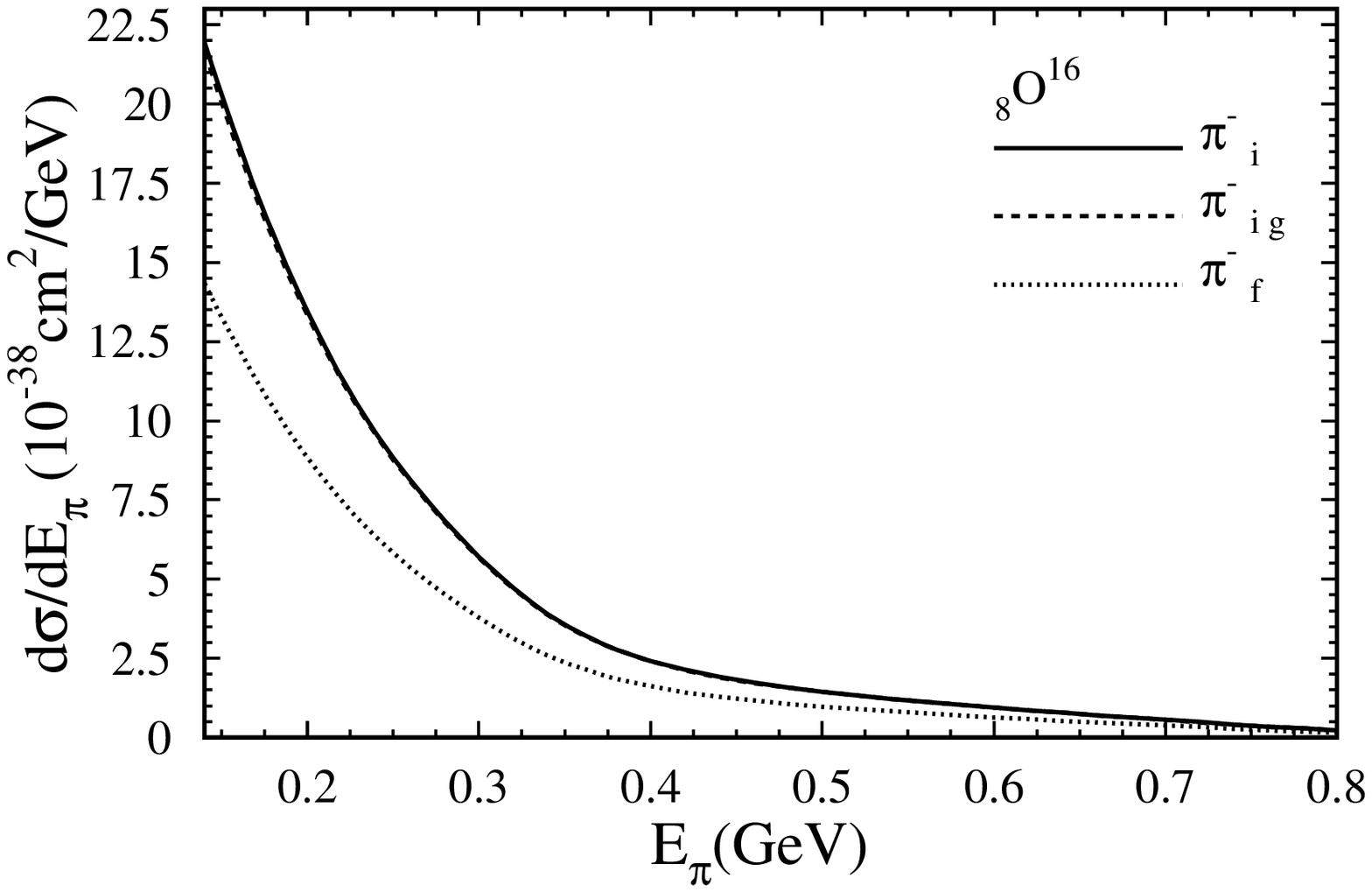,height=3.6in,angle=0}}
\caption{Pion energy distribution for negatively charged pions produced on oxygen targets. The solid, dashed and dotted lines represent respectively the pion energy distribution without any nuclear correction, including only the Pauli production factor $g$ and including all nuclear corrections.}\label{om}
\end{figure}

\begin{figure}
\begin{minipage}[b]{.49\linewidth}
\centering\epsfig{file=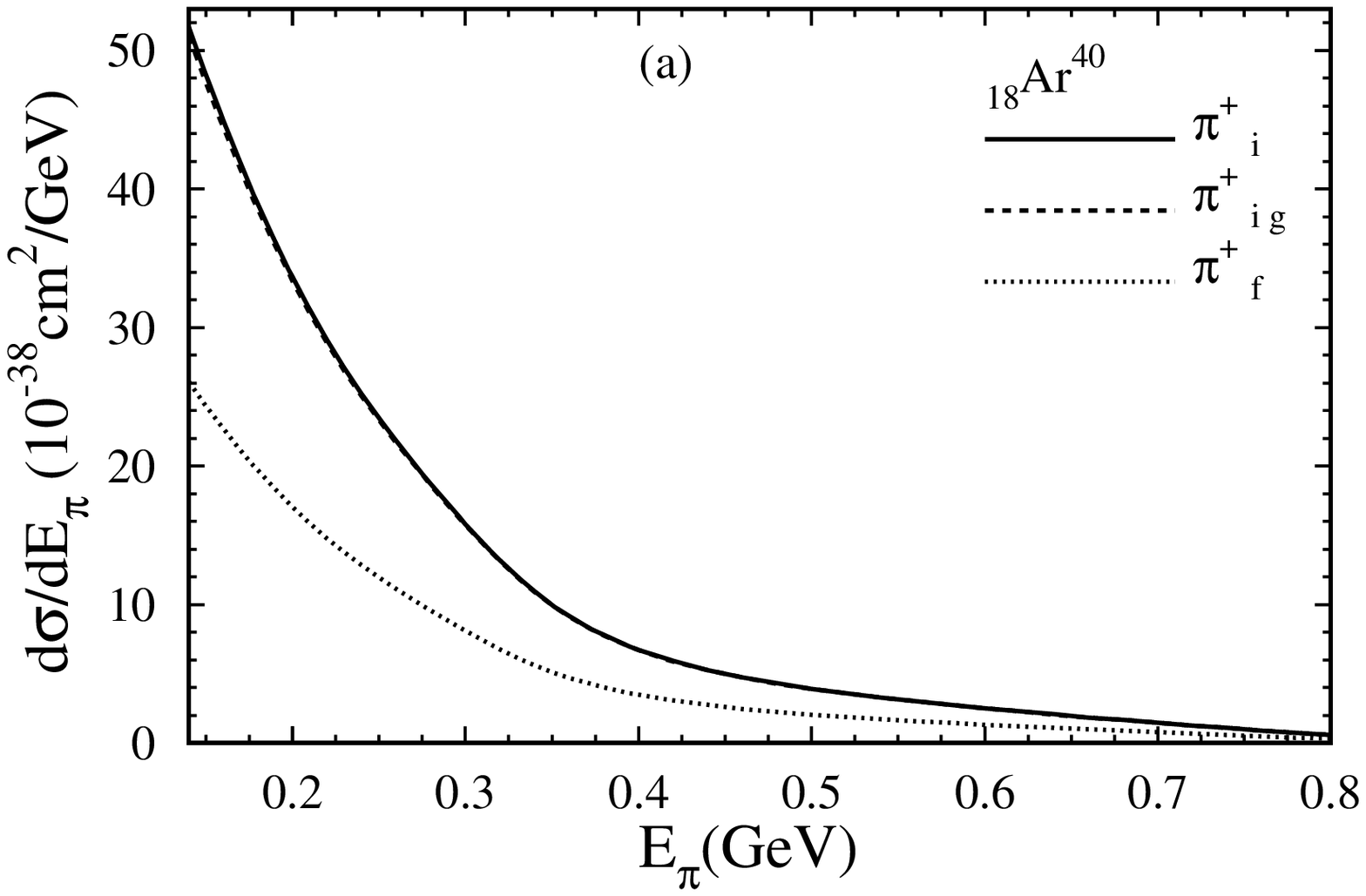,height=3.6in,width=\linewidth}
\end{minipage}
\begin{minipage}[b]{.49\linewidth}
\centering\epsfig{file=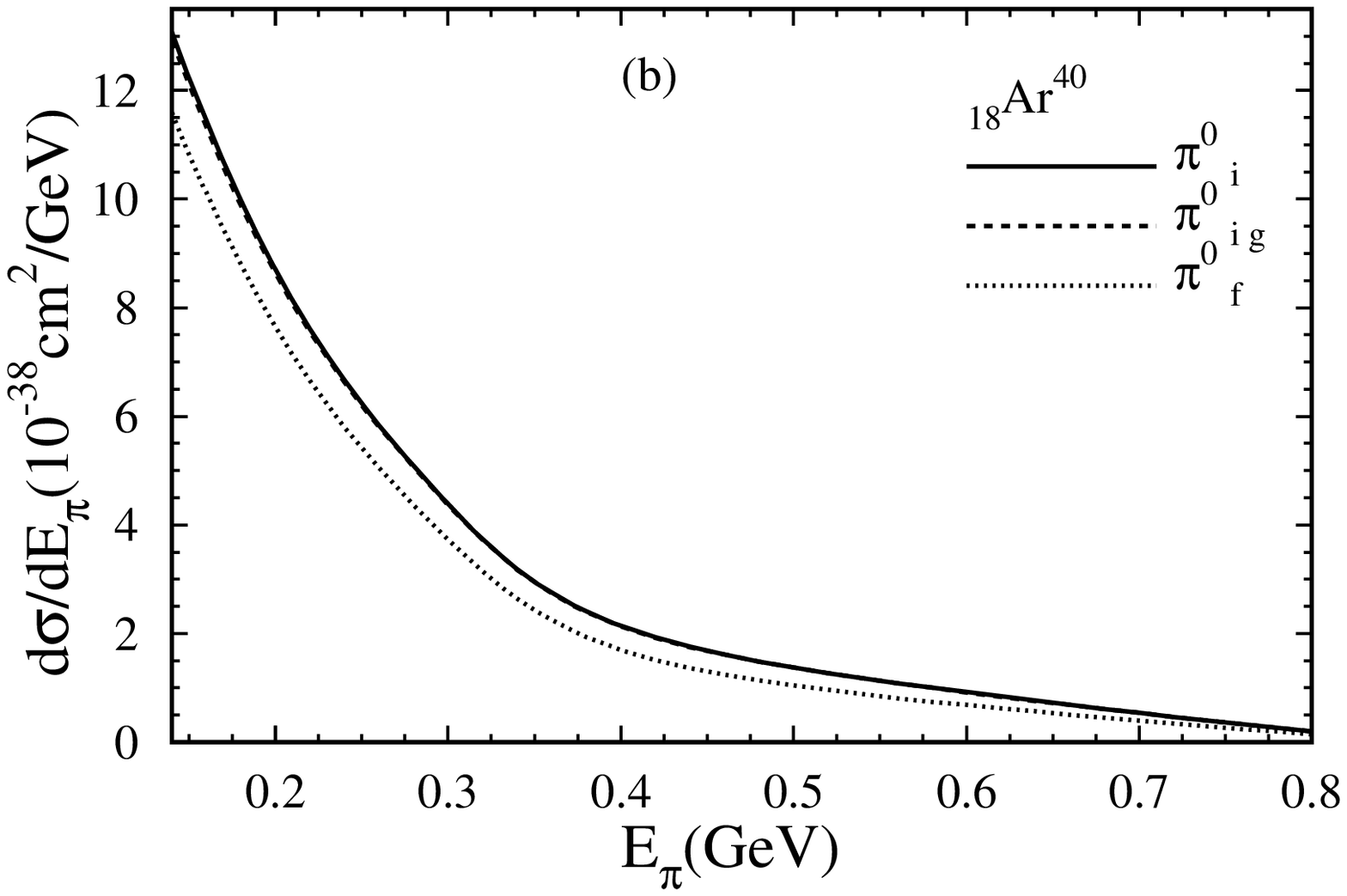,height=3.6in,width=\linewidth}
\end{minipage}
\vspace{-1.0cm}
\caption{Pion energy distribution for (a) positively charged pion and (b) neutrally pion produced on argon targets. The solid, dashed and dotted lines represent respectively the pion energy distribution without any nuclear correction, including only the Pauli production factor $g$ and including all nuclear corrections.}\label{ar0}
\end{figure}

\begin{figure}
\centerline{\psfig{figure=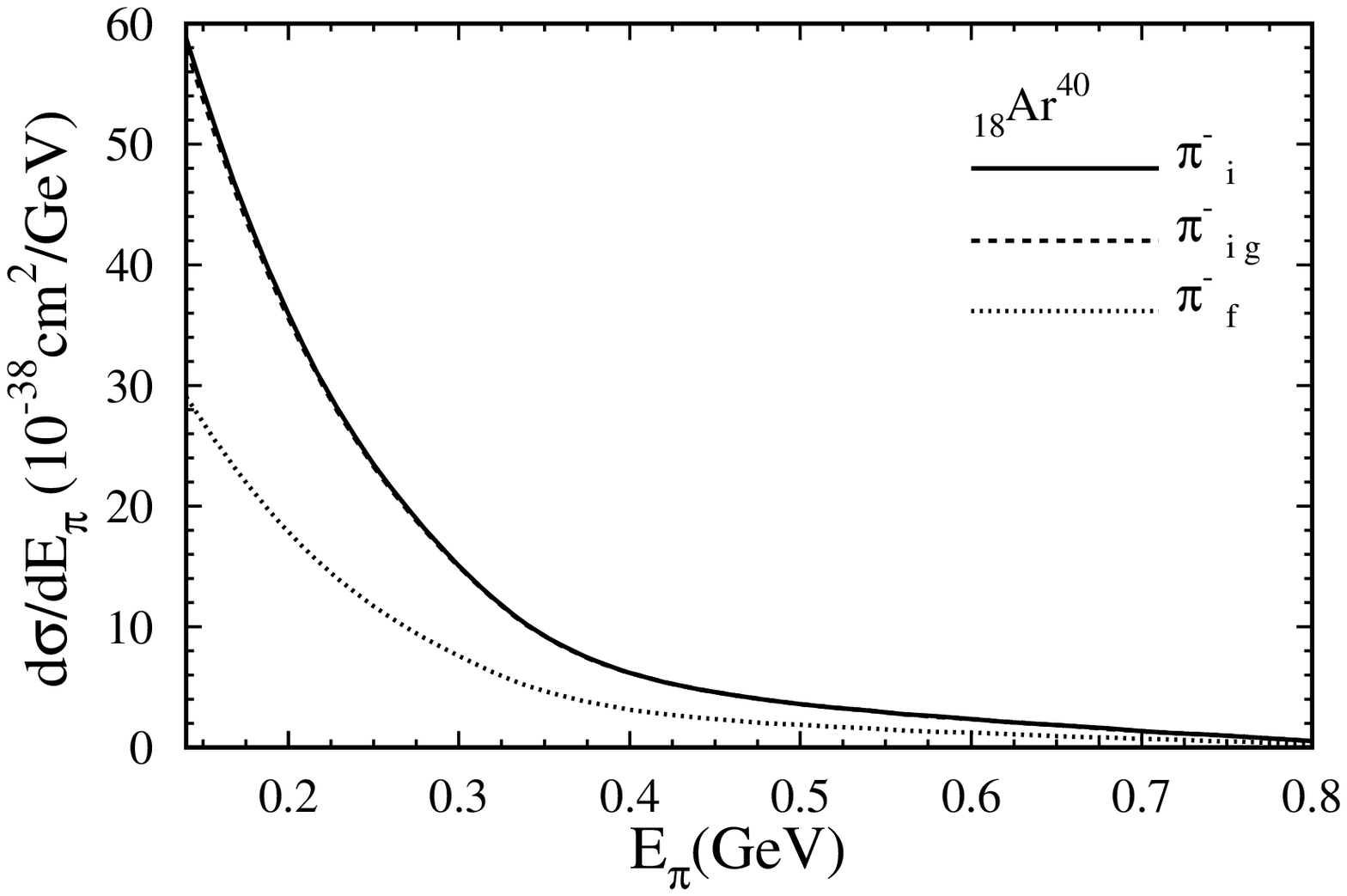,height=3.6in,angle=0}}
\caption{Pion energy distribution for negatively charged pions produced on argon targets. The solid, dashed and dotted lines represent respectively the pion energy distribution without any nuclear correction, including only the Pauli production factor $g$ and including all nuclear corrections.}\label{arm}
\end{figure}

\begin{figure}
\begin{minipage}[b]{.49\linewidth}
\centering\epsfig{file=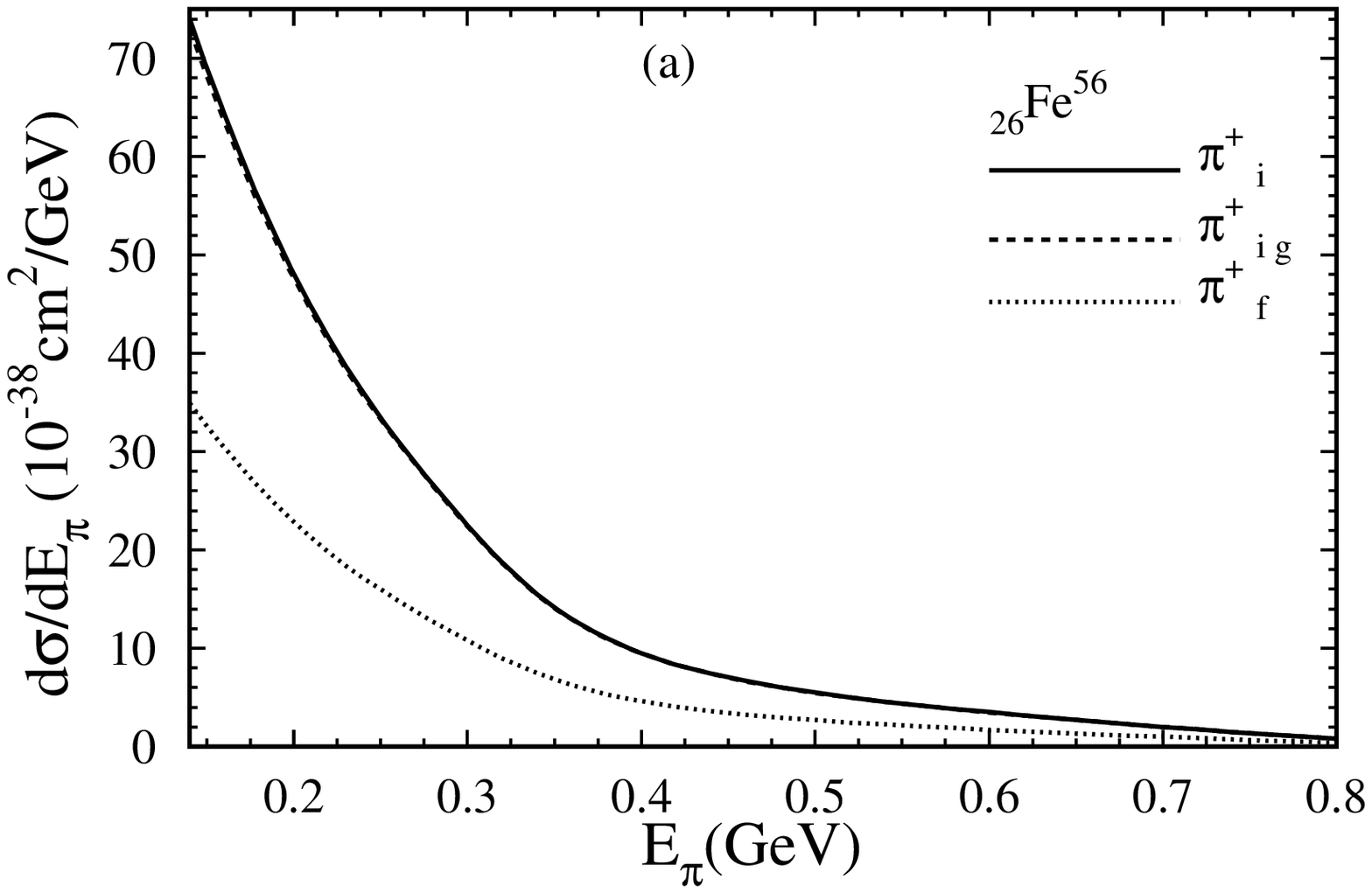,height=3.6in,width=\linewidth}
\end{minipage}
\begin{minipage}[b]{.49\linewidth}
\centering\epsfig{file=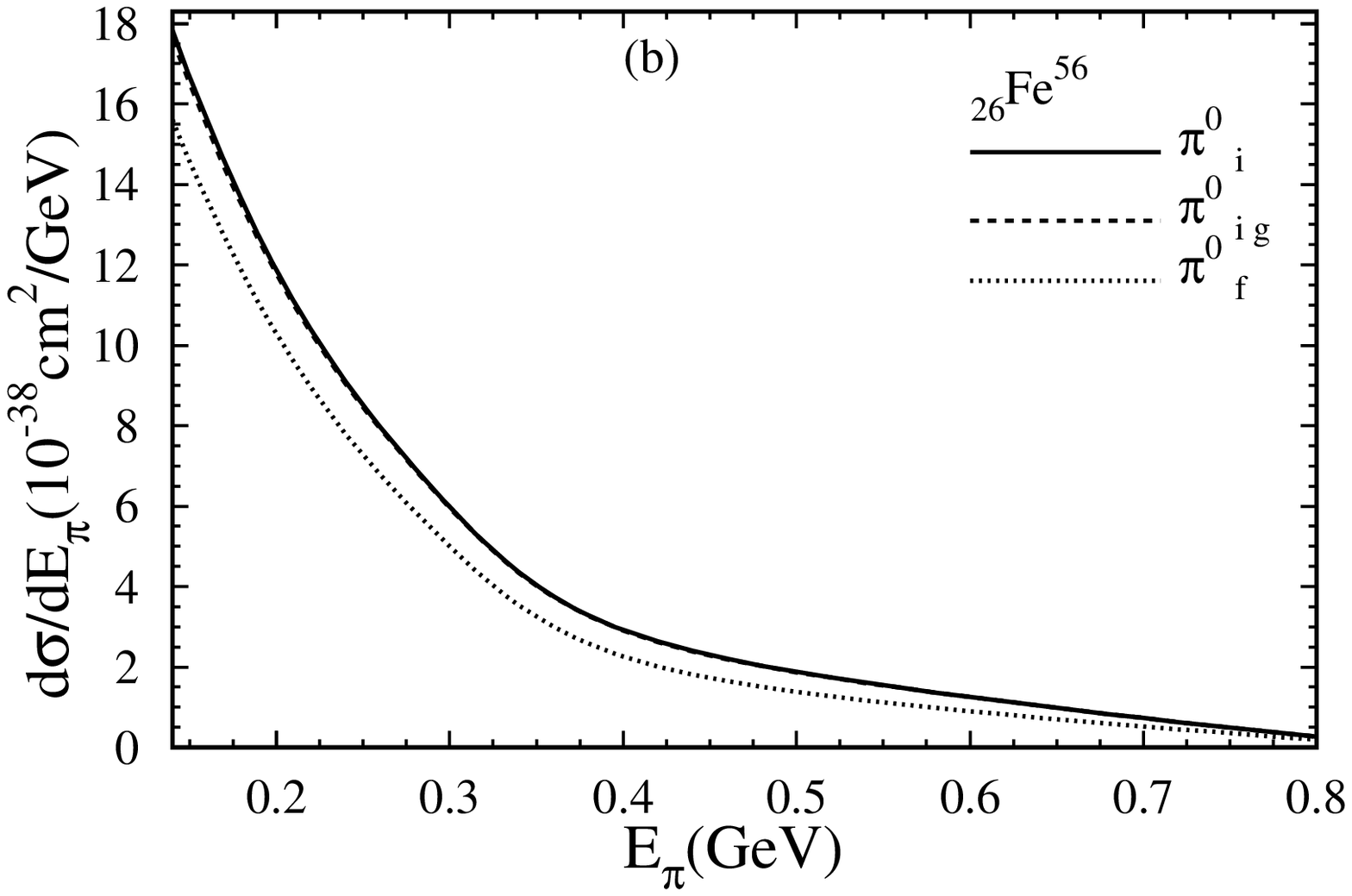,height=3.6in,width=\linewidth}
\end{minipage}
\vspace{-1.0cm}
\caption{Pion energy distribution for (a) positively charged pion and (b) neutral pion produced on iron targets. The solid, dashed and dotted lines represent respectively the pion energy distribution without any nuclear correction, including only the Pauli production factor $g$ and including all nuclear corrections.}\label{fe0}
\end{figure}

\clearpage


\begin{references}

\bibitem{Fukuda} 
Y. Fukuda {\it et al.}, Phys. Lett. B {\bf 335}, 237 (1994); Phys. Lett. B {\bf 433}, 9 (1998); Phys. Lett. B {\bf 436}, 33 (1998); Phys. Rev. Lett. {\bf 81}, 1562 (1998).

\bibitem{Casper} 
D. Casper {\it et al.}, Phys. Rev. Lett. {\bf 66}, 2561 (1991); R. Becker-Szendy {\it et al.}, Phys. Rev. D {\bf 46}, 3720 (1992).

\bibitem{Hirata} 
K. S. Hirata {\it et al.}, Phys. Lett. B {\bf 205}, 416 (1988); Phys. Lett. B {\bf 280}, 146 (1992).

\bibitem{K2k}
K. Nishikawa {\it et. al.} Nucl. Phys. (Proc. Suppl.) B {\bf 59} 289 (1997).

\bibitem{Itow}
Y.Itow {\it et al.}, eprint-archive hep-ex/0106019.

\bibitem{Minos}
P. G. Harris, Nucl. Phys. Proc. Suppl. B {\bf 85} 113 (2000); J. Schneps, Nucl. Phys. Proc. Suppl. B {\bf 87}, 189 (2000). 

\bibitem{Opera}
H. Shibuya {\it et al.}, LONG-LOI 8/97; K. Kodama {\it et al.}, CERN/SPSC 98-25; CERN/SPSC99-20; A. G. Gocco, Nucl. Phys. Proc. Suppl. B {\bf 85}, 125 (2000). 
\bibitem{Icanoe}
P. Assimaopoulos {\it et al.}, CERN/SPSC 99-25.

\bibitem{Alb}
C. H. Albright {\it et al.}, FERMILAB-FN-692, eprint-archive hep-ex/0008064; B. Autin, A. Blonde and J. Ellis, CERN-99-02.

\bibitem{Hall}
Lawrence J. Hall and Hitoshi Murayama, Phys. Lett. B {\bf 463} (1999). 

\bibitem{Apollonio} 
M. Apollonio {\it et al.}, Phys. Lett. B {\bf 420}, 397 (1998);  B {\bf 466}, 415 (1999).

\bibitem{Yu}
E. A. Paschos, L. Pasquali and J.Y. Yu, Nucl. Phys. B {\bf 588}, 263 (2000).

\bibitem{Derman}
E. Derman, Phys. Rew. D {\bf 7}, 2755 (1973).

\bibitem{Particle Data}
Review of Particle Physics, Eur. Phys. J. C {\bf 15}, 73 (2000).

\bibitem{Albright}
C. H. Albright and C. Jarlskog, Nucl. Phys. B {\bf84}, 467 (1975).

\bibitem{Lai}
H. L. Lai {\it et al.}, eprint-archive hep-ph/9903282.

\bibitem{Llewellyn}
C. H. Llewellyn Smith, Phys. Rep. {\bf 3}, 261 (1972).

\bibitem{Kum}
M. Hirai, S. Kumano and M. Miyama, eprint-archive hep-ph/0103208.

\bibitem{Esk}
K.J. Eskola, V.J. Kolhinen and C.A. Salgado,  Eur. Phys. J. C {\bf 9}, 61 (1999).

\bibitem{Esk1}
K.J. Eskola, V.J. Kolhinen and P.V. Ruuskanen, Nucl. Phys. B {\bf 535}, 351 (1998); eprint-archive hep-ph/9906484. 

\bibitem{Lov}
J. L$\phi$vseth, Phys. Lett. {\bf 5}, 199 (1963); Nuovo Cimento {\bf 57}, 382 (1968). 

\bibitem{Yao}
Y. P. Yao, Phys. Rev. {\bf 176}, 1680 (1968). 

\bibitem{Battistoni}
G. Battistoni {\it et al.}, eprint-archive hep-ph/9801426.

\bibitem{Bell}
J. S. Bell and C. H. Llewellyn Smith, Nucl. Phys. B {\bf 28}, 317 (1971).

\bibitem{Singh}
S. K. Singh and E. Oset, Nucl. Phys. A {\bf 542}, 587 (1992).

\bibitem{Bohr}
A. Bohr and B. R. Mottelson, Nuclear structure vol. {\bf I} (1969). 

\bibitem{Barish}
S. J. Barish {\it et al.}, Phys. Rev. D {\bf 19}, 2521 (1979).

\bibitem{Dolly}
D. C. Dolly {\it et al.}, Z. Phys. C {\bf 2}, 187 (1979).

\bibitem{Baker}
N. J. Baker {\it et al.}, Phys. Rev. D {\bf 25}, 617 (1982).

\bibitem{Auc}
P. S. Auchincloss {\it et al.}, Z. Phys. C {\bf 48}, 411 (1990).

\bibitem{Seligman}
W. Seligman, Ph. D. Thesis, Nevis Report 292 (1996).

\bibitem{Mac}
D. B. MacFarlane {\it et al.}, Z. Phys. C {\bf 26}, 1 (1984).

\bibitem{Berge}
P. Berge {\it et al.}, Z. Phys. C {\bf 35}, 433 (1987).

\bibitem{All}
J. V. Allaby {\it et al.}, Z. Phys. C {\bf 38}, 403 (1988).

\bibitem{Bal}
C. Baltay {\it et al.}, Phys. Rew. Lett. {\bf 44}, 916 (1980).

\bibitem{Cam}
S. Campolillo {\it et al.}, Phys. Lett. B {\bf 84} ,281 (1979).

\bibitem{Mor}
J. Morfin {\it et al.}, Phys. Lett. B {\bf 104}, 235 (1981).

\bibitem{Vov}
A. S. Vovenko {\it et al.}, Sov. J. Nucl. Phys. {\bf 30}, 527 (1979).

\bibitem{Bar}
D. S. Baranov {\it et al.}, Phys. Lett. B {\bf 81}, 225 (1979).

\bibitem{Asl}
E. Aslanides {\it et al.}, eprint-archive astro-ph/9907432.

\bibitem{Resv}
L. K. Resvanis, Nuc. Phys. Proc. Suppl. B {\bf 87}, 448 (2000).

\bibitem{And}
E. Andres {\it et al.}, Nuc. Phys. Proc. Suppl. B {\bf 91}, 423 (2000).

\bibitem{Zha}
Zh.- A. Dzhilkibaev, astro-ph/0105269.

\bibitem{Adler}
S. L. Adler, S. Nussinov and E. A. Paschos, Phys. Rev. D {\bf 9}, 2125 (1974).

\bibitem{NGS}
The neutrino flux $\nu_\mu$ is taken from the http://www.cern.ch/NGS.

\bibitem{Gago}
A. M. Gago {\it et al.}, eprint-archive hep-ph/0010092.

\bibitem{Gon}
M. C. Gonzalez-Garcia {\it et al.}, eprint-archive hep-ph/0009350.\label{Gon}

\bibitem{Rubbia}
A. Rubbia {\it et al.}, eprint-archive hep-ex/0008071.

\bibitem{Oset}
S. K. Singh, M. J. Vincente Vacas and E. Oset, Phys. Lett. B {\bf 416}, 23 (1998).

\bibitem{Bari}
S. J. Barish {\it et al.}, Phys. Rev. D {\bf 16}, 3103 (1977).

\bibitem{Bonetti}
S. Bonetti {\it et al.}, Nuovo Cimento A {\bf 38}, 260 (1977).

\bibitem{Belikov}
S. V. Belikov {\it et al.}, Z. Phys. A {\bf 320}, 625 (1985).

\end{references}
\end{document}